\documentclass[a4paper,12pt]{article}

\pdfoutput=1
\usepackage{amsmath}
\usepackage{amssymb}
\usepackage{amsfonts}
\usepackage{mathrsfs}
\usepackage{bbm}
\usepackage{graphicx,subfigure,booktabs}
\usepackage[numbers,sort&compress]{natbib}
\usepackage{verbatim}
\usepackage{color}
\usepackage{ulem}
\usepackage{setspace}
\usepackage{multirow}
\usepackage{url}
\usepackage{cancel}

\usepackage[utf8]{inputenc}

\usepackage[colorlinks,
linkcolor=black,
filecolor=black,
anchorcolor=black,
urlcolor=black,
citecolor=blue,
bookmarks=false,
]{hyperref}

\usepackage[hyphenbreaks]{breakurl}

\numberwithin{equation}{section}

\newlength{\dinwidth}
\newlength{\dinmargin}
\makeatletter
\newcommand{\thickhline}{%
	\noalign {\ifnum 0=`}\fi \hrule height 1pt
	\futurelet \reserved@a \@xhline
}
\makeatother
\allowdisplaybreaks

\begin{document}
	
\title{\vspace{-2.0cm} \bf CP asymmetry in \boldmath{$\tau\to K_S\pi\nu_\tau$} decays within the Standard Model and beyond}

\author{Feng-Zhi Chen\footnote{ggchan@mails.ccnu.edu.cn},
Xin-Qiang Li\footnote{xqli@mail.ccnu.edu.cn},
Ya-Dong Yang\footnote{yangyd@mail.ccnu.edu.cn},\,
and
Xin Zhang\footnote{xinzhang@mail.ccnu.edu.cn}\\[12pt]
\small Institute of Particle Physics and Key Laboratory of Quark and Lepton Physics~(MOE), \\
\small Central China Normal University, Wuhan, Hubei 430079, China}

\date{}
\maketitle
\vspace{-0.2cm}

\begin{abstract}
{\noindent}Motivated by the $2.8\sigma$ discrepancy observed between the BaBar measurement and the Standard Model prediction of the CP asymmetry in $\tau\to K_S\pi\nu_\tau$ decays, as well as the prospects of future measurements at Belle II, we revisit this observable in this paper. Firstly, we reproduce the known CP asymmetry due to $K^0 -\bar{K}^0$ mixing by means of the reciprocal basis, which is convenient when a $K_{S(L)}$ is involved in the final state. As the $K\pi$ tensor form factor plays a crucial role in generating a non-zero direct CP asymmetry that can arise only from the interference of vector and tensor operators, we then present a dispersive representation of this form factor, with its phase obtained in the context of chiral theory with resonances, which fulfills the requirements of unitarity and analyticity. Finally, the $\tau\to K_S\pi\nu_\tau$ decays are analyzed both within a model-independent low-energy effective theory framework and in a scalar leptoquark scenario. It is observed that the CP anomaly can be accommodated in the model-independent framework, even at the $1\sigma$ level, together with the constraint from the branching ratio of $\tau^-\to K_S\pi^-\nu_\tau$ decay; it can be, however, marginally reconciled only at the $2\sigma$ level, due to the specific relation between the scalar and tensor operators in the scalar leptoquark scenario. Once the combined constraints from the branching ratio and the decay spectrum of this decay are taken into account, these possibilities are however both excluded, even without exploiting further the stronger bounds from the (semi-)leptonic kaon decays under the assumption of lepton-flavour universality, as well as from the neutron electric dipole moment and $D-\bar{D}$ mixing under the assumption of $SU(2)$ invariance of the weak interactions. 
\end{abstract}

\newpage

\section{Introduction}
\label{sec:intro}

As the Kobayashi-Maskawa ansatz~\cite{Kobayashi:1973fv} for CP violation in the quark sector of the Standard Model (SM) is far too small to explain the observed baryon asymmetry of the universe~\cite{Sakharov:1967dj,Huet:1994jb,Cohen:1993nk,Riotto:1999yt}, we need to look for other sources of CP violation in different ways. This makes the CP-violating observables particularly interesting probes of new physics (NP) beyond the SM. In this respect, the hadronic decays of the $\tau$ lepton, besides serving as a clean laboratory for testing various low-energy aspects of the strong interaction~\cite{Pich:2013lsa,Davier:2005xq}, may also allow us to explore non-standard CP-violating interactions~\cite{Bigi:2012km,Bigi:2012kz,Kiers:2012fy}. 

In this paper, we shall focus on the CP asymmetry in $\tau\to K_S\pi\nu_\tau$ decays. After the initial null results from CLEO~\cite{Bonvicini:2001xz} and Belle~\cite{Bischofberger:2011pw}, a non-zero CP asymmetry was reported for the first time by the BaBar collaboration, with the result given by~\cite{BABAR:2011aa}
\begin{align}\label{eq:ACP_Exp}
A_{Q}=&\frac{\Gamma(\tau^+\to [\pi^+\pi^-]_{``K_S"}\pi^+\bar\nu_\tau)
	-\Gamma(\tau^-\to [\pi^+\pi^-]_{``K_S"}\pi^-\nu_\tau)}
{\Gamma(\tau^+\to[\pi^+\pi^-]_{``K_S"}\pi^+\bar\nu_\tau)
	+\Gamma(\tau^-\to[\pi^+\pi^-]_{``K_S"}\pi^-\nu_\tau)}\,\nonumber\\
=&(-0.36\pm0.23\pm0.11)\%\,,
\end{align}
where the first uncertainty is statistical and the second systematic. The subscript $``K_S"$ indicates that the intermediate $K_S$ is reconstructed in terms of a $\pi^+\pi^-$ final state with invariant mass around $M_K$ and at a decay time close to the $K_S$ lifetime. Within the SM, as there is no direct CP violation in hadronic $\tau$ decays at the tree level in weak interaction\footnote{The direct CP asymmetry generated by the second-order weak interaction is estimated to be of order $10^{-12}$, and can be therefore neglected safely~\cite{Delepine:2005tw}.}, this asymmetry arises solely from the CP violation in $K^0-\bar K^0$ mixing~\cite{Christenson:1964fg,Tanabashi:2018oca}, and is calculated to be~\cite{Bigi:2005ts,Grossman:2011zk}
\begin{align}\label{eq:ACP_SM1}
A_{CP}^{\text{SM}}=&\frac
{\int_{t_1}^{t_2}dt\,\left[\Gamma(K^0(t)\to\pi\pi)-\Gamma(\bar{K}^0(t)\to\pi\pi)\right]}
{\int_{t_1}^{t_2}dt\,\left[\Gamma(K^0(t)\to\pi\pi)+\Gamma(\bar{K}^0(t)\to\pi\pi)\right]} \nonumber\\
\approx&(3.32\pm0.06)\times10^{-3}\,,
\end{align}
where $K^0(t)$~($\bar{K}^0(t)$) denotes the time-evolved state identified at time $t=0$ as a pure $K^0$~($\bar{K}^0$), and the second line is obtained after neglecting the small correction from direct CP violation in $K\to \pi^+\pi^-$ decays and when $t_1\ll\tau_S$ and $\tau_S\ll t_2\ll\tau_L$ ($\tau_{S(L)}$ being the $K_{S(L)}$ lifetime). Such a CP asymmetry was predicted firstly by Bigi and Sanda~\cite{Bigi:2005ts} but with a sign mistake~\cite{Grossman:2011zk}. As emphasized by Grossman and Nir~\cite{Grossman:2011zk} (see also Ref.~\cite{Calderon:2007rg}), in the calculation of this CP asymmetry, the interference between the amplitudes of intermediate $K_S$ and $K_L$ is as important as the pure $K_S$ amplitude, and hence the measured CP asymmetry depends sensitively on the decay time interval over which it is integrated. After taking into account the $K_S\to \pi^+\pi^-$ decay-time dependence of the event selection efficiency, the BaBar collaboration obtained a multiplicative correction factor, $1.08\pm0.01$, for the CP asymmetry, with the resulting experimental data given by Eq.~\eqref{eq:ACP_Exp} and the corresponding SM prediction changed to $A_{CP}^{\text{SM}}=(0.36\pm0.01)\%$~\cite{BABAR:2011aa}. Thus, a $2.8\sigma$ discrepancy is observed between the BaBar measurement and the SM prediction and, if confirmed with a higher precision by Belle and/or Belle II~\cite{Kou:2018nap}, would be a clear NP signal.

Such a CP anomaly, together with the prospects of future measurements at Belle II~\cite{Kou:2018nap}, has motivated several studies of possible direct CP asymmetries in $\tau\to K_S\pi\nu_\tau$ decays due to non-standard interactions~\cite{Devi:2013gya,Dhargyal:2016kwp,Dhargyal:2016jgo,Cirigliano:2017tqn,Rendon:2019awg,Dighe:2019odu}. As argued in Refs.~\cite{Devi:2013gya,Cirigliano:2017tqn}, due to the lack of a relative strong phase, an explanation with scalar operators is already excluded\footnote{Although the interference of vector and scalar operators could still contribute to the CP asymmetry due to long-distance QED corrections~\cite{Antonelli:2013usa}, the scalar contribution is strongly suppressed and will be of little phenomenological relevance when the constraint from the $\tau\to K_S\pi\nu_\tau$ branching ratio is taken into account~\cite{Cirigliano:2017tqn}.}, and only the interference of vector and tensor operators can provide a possible strong phase difference, leaving new tensor interactions as the only potential NP explanation. Here, whether the tensor interaction is admissible to account for the anomaly or not depends crucially on the $K\pi$ tensor form factor. In Refs.~\cite{Devi:2013gya,GodinaNava:1995jb,Delepine:2006fv}, the tensor form factor was assumed to be a real constant, which is motivated by the analysis of $K_{\ell3}$ ($K\to \pi \ell \nu_\ell$ with $\ell=e,\,\mu$) data~\cite{Tanabashi:2018oca}, and the relative strong phase, being now just the phase of the vector form factor, was found to be large enough to produce a sizable CP asymmetry. This assumption was, however, pointed out to be incorrect by Cirigliano, Crivellin and Hoferichter~\cite{Cirigliano:2017tqn}. They demonstrated that, as the same spin-1 resonances contributing to the vector form factor will equivalently contribute to the tensor one, the crucial interference between vector and tensor phases is suppressed by at least two orders of magnitude due to Watson's final-state-interaction theorem~\cite{Watson:1954uc}, and the amount of CP asymmetry that a tensor operator can produce is, therefore, strongly suppressed~\cite{Cirigliano:2017tqn}. Such a conclusion is, however, based on the assumption that the inelastic contributions to the phases of vector and tensor form factors are of similar size but potentially opposite in sign~\cite{Cirigliano:2017tqn}.

In order to obtain sensible constraints on non-standard interactions from $\tau\to K_S\pi\nu_\tau$ decays, the exact distributions of the $K\pi$ form factors, including both their moduli and phases, as a function of $s=q^2$, the invariant mass squared of the $K\pi$ final state, are needed. For the vector and scalar form factors, either the Breit-Wigner parametrisations~\cite{Epifanov:2007rf,Paramesvaran:2009ec,Finkemeier:1995sr,Finkemeier:1996dh} or the dispersive representations~\cite{Jamin:2006tk,Jamin:2008qg,Boito:2008fq,Boito:2010me,Jamin:2000wn,Jamin:2001zq,Jamin:2006tj,Escribano:2014joa,Moussallam:2007qc,Bernard:2013jxa} can be used, with the relevant parameters determined via a successful fit to the measured $K\pi$ invariant mass spectrum~\cite{Epifanov:2007rf}. For the tensor form factor, however, there exists no experimental data that can guide us to construct it, and we have to rely on theory. While a $q^2$-independent tensor form factor or its normalization~\cite{Garces:2017jpz} can be derived from the leading-order chiral perturbation theory ($\chi$PT)~\cite{Weinberg:1978kz,Gasser:1983yg,Gasser:1984gg,Gasser:1984ux} with tensor sources~\cite{Cata:2007ns,Mateu:2007tr}, we have to get its $q^2$ dependence by solving numerically the dispersion relation~\cite{Miranda:2018cpf,Rendon:2019awg}, with its phase obtained in the context of chiral theory with resonances (R$\chi$T)~\cite{Ecker:1988te,Ecker:1989yg}. It should be mentioned that the tensor form factor given in Ref.~\cite{Miranda:2018cpf} is derived at the lowest chiral order (being $\mathcal{O}(p^4)$ in the chiral counting~\cite{Cata:2007ns}) of R$\chi$T and fails to satisfy the unitarity requirement, which could be compensated by including the contributions from the next-to-leading order (NLO) $\chi$PT Lagrangian with tensor sources. Although the spin-1 resonances can be described equivalently by vector or anti-symmetric tensor fields~\cite{Ecker:1988te,Ecker:1989yg}, it will be shown that the former is more convenient in describing the interactions of tensor currents with the resonances. The unitarity property will also be satisfied automatically when the NLO (being $\mathcal{O}(p^6)$ in the chiral counting~\cite{Cata:2007ns}) terms with the model~\textbf{II} prescription~\cite{Ecker:1989yg} are properly taken into account. In this paper, motivated by these two observations and following Refs.~\cite{Miranda:2018cpf,Rendon:2019awg}, we shall present an alternative dispersive representation of the tensor form factor, with its phase obtained in the context of R$\chi$T, which fulfills the requirements of unitarity and analyticity. 

Taking as inputs the three-times subtracted (for the vector form factor)~\cite{Boito:2008fq,Boito:2010me} and the coupled-channel (for the scalar form factor)~\cite{Jamin:2000wn,Jamin:2001zq,Jamin:2006tj} dispersive representations, together with our result of the tensor form factor, we shall then analyze the $\tau\to K_S\pi\nu_\tau$ decays both within a model-independent low-energy effective theory framework and in a scalar leptoquark~(LQ) scenario~\cite{Bauer:2015knc}. It will be shown that the CP anomaly can be accommodated in the model-independent framework, even at the $1\sigma$ level, together with the constraint from the branching ratio of $\tau^-\to K_S\pi^-\nu_\tau$ decay. In the LQ scenario, however, this anomaly can be marginally reconciled only at the $2\sigma$ level, due to the specific relation between the scalar and tensor operators. Once the combined constraints from the branching ratio and the decay spectrum of this decay are taken into account, these possibilities are however both excluded, even without exploiting further the stronger bounds from the (semi-)leptonic kaon decays~\cite{Gonzalez-Alonso:2016etj} under the assumption of lepton-flavour universality, as well as from the neutron electric dipole moment~(EDM) and $D-\bar{D}$ mixing under the assumption of $SU(2)$ invariance of the weak interactions~\cite{Cirigliano:2017tqn}. It is therefore quite difficult to explain such a CP anomaly within the frameworks considered here. 

Our paper is organized as follows. In section~\ref{sec:ACPSM}, we recalculate the CP asymmetry due to $K^0 -\bar{K}^0$ mixing by means of the reciprocal basis, and reproduce the result given in Ref.~\cite{Grossman:2011zk}. In section~\ref{sec:new physics}, the $\tau\to K_S\pi\nu_\tau$ decays are analyzed both within the model-independent framework and in the scalar LQ scenario. Section~\ref{sec:FFs} is devoted to the calculation of the $K\pi$ tensor form factor in the context of $\chi$PT  with tensor sources and R$\chi$T with the spin-1 resonances described by the conventional vector fields. Numerical results and discussions are then presented in section~\ref{sec:numerical analysis}. Our conclusions are finally made in section~\ref{sec:conclusion}. For convenience, all the input parameters used throughout this paper are collected in the appendix.

\section{CP asymmetry in \boldmath{$\tau\to K_S\pi\nu_\tau$} decays within the SM}
\label{sec:ACPSM}

\begin{figure}[ht]
	\centering
	\includegraphics[width=0.38\textwidth]{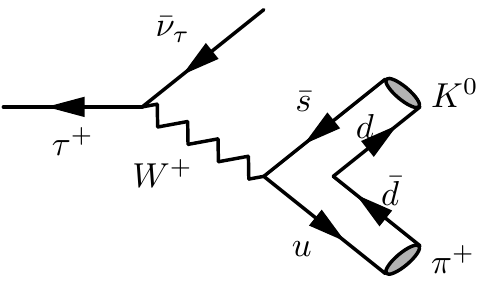}
	\hspace{0.45in}
	\includegraphics[width=0.38\textwidth]{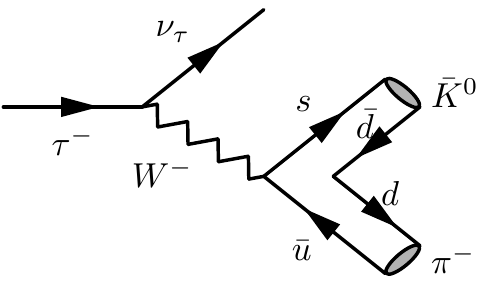}
	\caption{\label{fig:Feynman diagram} \small Tree-level Feynman diagrams for the decay $\tau^+\to K^0\pi^+\bar\nu_\tau$ (left) as well as its CP-conjugated mode $\tau^-\to \bar{K}^0\pi^-\nu_\tau$ (right) within the SM.}
\end{figure}

Before discussing the CP asymmetry in $\tau\to K_S\pi\nu_\tau$ decays within the SM, one should firstly notice that the $\tau^+$~($\tau^-$) decay produces initially a $K^0$~($\bar{K}^0$) state due to the $\Delta S=\Delta Q$ rule, and the relevant Feynman diagrams at the tree level in weak interaction are shown in Fig.~\ref{fig:Feynman diagram}. As the involved CKM matrix element $V_{us}$ is real and the strong phase must be the same in these two CP-related processes, the transition amplitudes within the SM should satisfy
\begin{align}\label{eq:amp_sm}
\mathcal{A}(\tau^+\to K^0\pi^+\bar\nu_\tau)=\mathcal{A}(\tau^-\to \bar{K}^0\pi^-\nu_\tau)\,.
\end{align}
Due to the $K^0-\bar{K}^0$ mixing, on the other hand, the experimentally reconstructed kaons are the mass ($|K_S\rangle$ and $|K_L\rangle$) rather than the flavour ($|K^0\rangle$ and $|\bar{K}^0\rangle$) eigenstates, which, in the absence of CP violation in the system, are related to each other via~\cite{Silva:2004gz}
\begin{align}\label{eq:mixing_no_cp}
|K_{S,L}\rangle=\frac{1}{\sqrt{2}}\left(|K^0\rangle \pm e^{i\zeta}|\bar{K}^0\rangle\right)\,,
\end{align}
where $\zeta$ is the spurious phase brought about by the CP transformation, $\mathcal{CP}|K^0\rangle=e^{i\zeta}|\bar{K}^0\rangle$~\cite{Silva:2004gz}. Then, one can get
\begin{align}\label{eq:width_no_cp}
\Gamma(\tau^+\to K_{S,L}\pi^+\bar\nu_\tau)=\frac{1}{2}\,\Gamma(\tau^+\to K^0\pi^+\bar\nu_\tau)\,,\nonumber\\[0.2cm]
\Gamma(\tau^-\to K_{S,L}\pi^-\nu_\tau)=\frac{1}{2}\,\Gamma(\tau^-\to \bar{K}^0\pi^-\nu_\tau)\,,
\end{align}
which, after taken together with Eq.~\eqref{eq:amp_sm}, would indicate that there exists no CP asymmetry in $\tau\to K_S\pi\nu_\tau$ decays within the SM. Furthermore, the contribution from second-order weak interaction is estimated to be of order $10^{-12}$, and can be therefore neglected safely~\cite{Delepine:2005tw}.

Once the CP violation in $K^0-\bar{K}^0$ mixing~\cite{Christenson:1964fg,Tanabashi:2018oca} is included, however, a non-zero CP asymmetry would appear in $\tau\to K_S\pi\nu_\tau$ decays, as elaborated in Refs.~\cite{Bigi:2005ts,Grossman:2011zk}. In order to see this clearly, we shall follow the convention specified in Ref.~\cite{Silva:2004gz} and recalculate this asymmetry by means of the reciprocal basis~\cite{Sachs:1963zz,Enz:1965tr,Wolfenstein:1970wb,Beuthe:1997fu,AlvarezGaume:1998yr,Branco:1999fs,Silva:2000db,Silva:2004gz}. In the presence of CP violation but with the CPT invariance still assumed, the two mass eigenkets are now given by~\cite{Silva:2004gz}
\begin{align}\label{eq:mixing_cp}
|K_{S,L}\rangle=p\,|K^0\rangle \pm q\,|\bar{K}^0\rangle\,,
\end{align}
with the normalization $|p|^2+|q|^2=1$, and the corresponding mass eigenbras read~\cite{Silva:2004gz}
\begin{align}\label{eq:reciprocal}
\langle \tilde K_{S,L} |=\frac{1}{2}\left(p^{-1}\langle K^0| \pm q^{-1}\langle \bar{K}^0|\right)\,,
\end{align}
which form the so-called reciprocal basis ($\langle \tilde K_{S} |$,  $\langle \tilde K_{L}|$) that is featured by both the orthornormality and the completeness conditions~\cite{Silva:2004gz}:
\begin{align}\label{eq:relation}
& \langle \tilde K_{S} |K_{S}\rangle=\langle \tilde K_{L} |K_{L}\rangle=1\,,\quad
\langle \tilde K_{S} |K_{L}\rangle=\langle \tilde K_{L} |K_{S}\rangle=0\,,\nonumber\\[0.2cm]
& \hspace{2.5cm} |K_{S}\rangle\langle \tilde K_{S} |+|K_{L}\rangle\langle \tilde K_{L} |=1\,.
\end{align}
Then, the time-evolution operator for the $ K^0-\bar K^0 $ system is determined by
\begin{equation}
\text{exp}(-i\mbox{\boldmath $H$}t)=e^{-i\mu_{S}t}|K_S\rangle\langle \tilde K_S|+e^{-i\mu_{L}t}|K_L\rangle\langle \tilde K_L|,
\end{equation}
where $\mu_S=M_S-i/2\,\Gamma_S$ and $\mu_L=M_L-i/2\,\Gamma_L$ are the two eigenvalues of the $2\times 2$ effective Hamiltonian $\mbox{\boldmath $H$}$ used to describe the $K^0-\bar{K}^0$ mixing.

The intermediate $K_S$ in $\tau\to K_S\pi\nu_\tau$ decays is not directly observed in experiment, but reconstructed via a $\pi^+\pi^-$ final state with $M_{\pi\pi}\approx M_K$ and a time difference $t\approx \tau_S$ between the $\tau$ and the $K$ decay~\cite{Grossman:2011zk}. However, as the CP symmetry is violated in $K^0-\bar{K}^0$ mixing~\cite{Christenson:1964fg,Tanabashi:2018oca},  the final state $\pi^+\pi^-$ can be obtained not only from $K_S$ but also from $K_L$, for long decay times of kaons.  As a consequence, the complete amplitude for the process $\tau^+\to [\pi^+\pi^-]\pi^+\bar\nu_\tau$ involves the amplitude for the initial $\tau^+$ decay into the intermediate state $K_{S,L}\pi^{+}\bar \nu_{\tau}$, the time-evolution amplitude for this state, and finally the amplitude for the decay into $[\pi^+\pi^-]\pi^{+}\bar \nu_{\tau}$. Suppressing the reference to $\pi^{+}\bar \nu_{\tau}$, we can therefore write the complete amplitude as~\cite{Silva:2000db}
\begin{align}\label{eq:taupmdecay}
\mathcal{A}(\tau^+\to K_{S,L}\to\pi^+\pi^-)&=\langle \pi^+\pi^-|T|K_S\rangle e^{-i\mu_St}\langle \tilde K_S|T|\tau^+\rangle+\langle \pi^+\pi^-|T|K_L\rangle e^{-i\mu_Lt}\langle \tilde K_L|T|\tau^+\rangle\,\nonumber\\[0.2cm]
&\hspace{-1.8cm}=\frac{1}{2p}\langle \pi^+\pi^-|T|K_S\rangle e^{-i\mu_St}\langle K^0|T|\tau^+\rangle+\frac{1}{2p}\langle \pi^+\pi^-|T|K_L\rangle e^{-i\mu_Lt}\langle K^0|T|\tau^+\rangle\,,
\end{align}
in which Eq.~\eqref{eq:reciprocal} and the $\Delta S=\Delta Q$ rule have been used to obtain the second line. Then, the complete time-dependent decay width for $\tau^+\to [\pi^+\pi^-]\pi^+\bar\nu_\tau$ can be written as\footnote{Our expression of the decay width is slightly different from that given in Ref.~\cite{Grossman:2011zk}, because the latter corresponds to the time-dependent decay width of $K^0$ decaying into the $2\pi$ system with isospin $I=0$, which contains both the $\pi^+\pi^-$ and $\pi^0\pi^0$ components, $\langle 2\pi,I=0|=\sqrt{\frac{2}{3}}\langle \pi^+\pi^-|-\sqrt{\frac{1}{3}}\langle \pi^0\pi^0|$. As the intermediate $K_S$ is reconstructed via the $\pi^+\pi^-$ final state in experiment~\cite{BABAR:2011aa}, one has to use $\eta_{+-}$ instead of $\epsilon=\frac{\langle 2\pi,I=0|T|K_L\rangle}{\langle 2\pi,I=0|T|K_S\rangle}$.}
\begin{align}
\Gamma(\tau^+\to [\pi^+\pi^-]\pi^+\bar\nu_\tau)
&=|\langle K^0| T|\tau^+\rangle|^2\, \nonumber\\[0.1cm]
&\hspace{-0.6cm}\times \frac{|\langle \pi^+\pi^-| T| K_S\rangle|^2}{4| p|^2}\Big[e^{-\Gamma_St}\!+\!|\eta_{+-}|^2\,e^{-\Gamma_Lt}\!+\!2|\eta_{+-}|\,e^{-\Gamma t}\cos(\phi_{+-}\!-\!\Delta m t) \Big]\,\label{eq:br}\\[0.2cm]
&=\Gamma(\tau^+\to K^0)\,\Gamma(K^0(t)\to\pi^+\pi^-)\,,\label{eq:evolution1}
\end{align}
where $\Gamma=\frac{\Gamma_L+\Gamma_S}{2}$ and $\Delta m=M_L-M_S$ denote respectively the average width and the mass difference of the $K^0-\bar K^0$ system, while the CP-violating amplitude ratio $\eta_{+-}$ is defined as
\begin{equation}
\eta_{+-}=\frac{\langle \pi^+\pi^-| T| K_L\rangle}{\langle \pi^+\pi^-| T| K_S\rangle}\,,
\end{equation}
with $|\eta_{+-}|=(2.232\pm 0.011)\times 10^{-3} $ and $ \phi_{+-}=(43.51\pm 0.05)^\circ $~\cite{Tanabashi:2018oca}. From Eq.~\eqref{eq:br} and the corresponding decay width for the CP-conjugated process $\tau^-\to [\pi^+\pi^-]\pi^-\nu_\tau$ (obtained from Eq.~\eqref{eq:br} by replacing $p$ and $+2|\eta_{+-}|$ with $q$ and $-2|\eta_{+-}|$, respectively), one can see that, for the sum of the two decay widths, both the interference (the last)  and the pure $K_L$ (the second) term are suppressed compared to the pure $K_S$ (the first term in the bracket) contribution. For the difference of the two decay widths, however, the interference between the amplitudes of $K_S$ and $K_L$ is found to be as important as the pure $K_S$ amplitude~\cite{Grossman:2011zk}.

The time dependence of the decay width in Eq.~\eqref{eq:br} makes the measurement of the CP asymmetry in $\tau\to K_S\pi\nu_\tau$ decays sensitive to the experimental cuts~\cite{Grossman:2011zk}: one has to take into account not only the efficiency as a function of the kaon decay time, but also the kaon energy in the laboratory frame to account for the time dilation. Parametrizing all these experiment-dependent effects by a function $F(t)$~\cite{Grossman:2011zk},  one can write the total CP asymmetry as~\cite{Devi:2013gya}
\begin{align}
A_{CP}(t_1,t_2)&\!=\!
\frac{\Gamma_{\tau^+}\int_{t_1}^{t_2}dt\,F(t)\,\Gamma(K^0(t)\to \pi^+\pi^-)-\Gamma_{\tau^-}\int_{t_1}^{t_2}dt\,F(t)\,\Gamma({\bar K^0}(t)\to\pi^+\pi^-)}
{\Gamma_{\tau^+}\int_{t_1}^{t_2}dt\,F(t)\,\Gamma(K^0(t)\to\pi^+\pi^-)+\Gamma_{\tau^-}\int_{t_1}^{t_2}dt\,F(t)\,
\Gamma({\bar K^0}(t)\to\pi^+\pi^-)}\nonumber\\[0.2cm]
&=\frac{A_{CP}^\tau+ A_{CP}^{K}(t_1,t_2)}{1+A_{CP}^\tau A_{CP}^{K}(t_1,t_2)}\,,\label{eq:ACP}
\end{align}
where $\Gamma_{\tau^\pm}=\Gamma(\tau^\pm\to K^0(\bar K^0)\pi^\pm\bar\nu_\tau(\nu_\tau))$ instead of $\Gamma(\tau^\pm\to K_S\pi^\pm\bar\nu_\tau(\nu_\tau))$ as defined in Ref.~\cite{Devi:2013gya}, while $A_{CP}^\tau$ and $A_{CP}^{K}(t_1,t_2)$ are defined, respectively, as
\begin{align}
A_{CP}^\tau&\equiv
\frac{\Gamma(\tau^+\to K^0\pi^+\bar \nu_\tau)-\Gamma(\tau^-\to \bar K^0\pi^-\nu_\tau)}
{\Gamma(\tau^+\to K^0\pi^+\bar \nu_\tau)+\Gamma(\tau^-\to \bar K^0\pi^-\nu_\tau)}\,,\label{eq:ACPTau}\\[0.2cm]
A_{CP}^{K}(t_1,t_2)&\equiv\frac
{\int_{t_1}^{t_2}dt\,F(t)\left[\Gamma(K^0(t)\to\pi^+\pi^-)-\Gamma(\bar{K}^0(t)\to\pi^+\pi^-)\right]}
{\int_{t_1}^{t_2}dt\,F(t)\left[\Gamma(K^0(t)\to\pi^+\pi^-)+\Gamma(\bar{K}^0(t)\to\pi^+\pi^-)\right]}\,.
\label{eq:ACPK}
\end{align}
Here $A_{CP}^\tau$ denotes the direct CP asymmetry induced by potential NP dynamics, while $A_{CP}^{K}(t_1,t_2)$ represents the indirect CP asymmetry originating from the $K^0-\bar{K}^0$ mixing. Within the SM, $\Gamma_{\tau^+}=\Gamma_{\tau^-}$, implying that $A_{CP}^\tau=0$, and only $A_{CP}^{K}(t_1,t_2)$ makes a non-zero contribution~\cite{Bigi:2005ts,Grossman:2011zk}.

Plugging into Eq.~\eqref{eq:ACPK} the expressions of the time-dependent decay widths $\Gamma(K^0(t)\to\pi^+\pi^-)$ and $\Gamma(\bar{K}^0(t)\to\pi^+\pi^-)$ (see Eqs.~\eqref{eq:br} and \eqref{eq:evolution1}) and neglecting all the terms suppressed by $\mathcal{O}(|\eta_{+-}|^2)$, one can finally reproduce the result given in Ref.~\cite{Grossman:2011zk}\footnote{If the $K_L$ contributions to the decay width in Eq.~\eqref{eq:br} were neglected, on the other hand, one would obtain a result which is of the same magnitude but opposite in sign with the prediction made in Ref.~\cite{Bigi:2005ts}.},
\begin{equation}\label{eq:our}
A_{CP}^{\text{SM}}\approx +2\Re e(\epsilon)=3.32\times10^{-3}\qquad \text{for \quad $t_1 \ll\Gamma_S^{-1}$ and $\Gamma_S^{-1}\ll t_2 \ll\Gamma_L^{-1}$} \,,
\end{equation}
in which the approximations with $|\eta_{+-}|\approx \frac{2\Re e(\epsilon)}{\sqrt{2}}$, $\phi_{+-}\approx 45^\circ$, $\Gamma\approx\frac{\Gamma_S}{2}$, and $\Delta m\approx\frac{\Gamma_S}{2}$~\cite{Branco:1999fs}, as well as a particular efficiency function $F(t)$~\cite{Grossman:2011zk},
\begin{align}
F(t)=\left\{
\begin{aligned}
1 & & t_1<t< t_2\\
0 & & \text{otherwise}
\end{aligned}
\right.
\end{align}
have been used. The SM CP asymmetry in Eq.~\eqref{eq:our}, after multiplied by the correction factor $1.08\pm0.01$~\cite{BABAR:2011aa}, is then changed to be $(0.36\pm0.01)\%$, as obtained by the BaBar collaboration~\cite{BABAR:2011aa}.

\section{\boldmath{$\tau\to K_S\pi\nu_\tau$} decays in the presence of NP dynamics}
\label{sec:new physics}

When NP dynamics beyond the SM are present, a non-zero direct CP asymmetry $A_{CP}^\tau$ can exist and hence contributes to the total CP asymmetry $A_{CP}(t_1,t_2)$. As neither the pseudo-scalar nor the axial-vector interaction can produce the $K\pi$ final state due to the parity conservation in strong interaction, and the scalar interaction cannot create a non-zero direct CP asymmetry due to the lack of a relative strong phase, we are left only with the tensor interaction as a possible mechanism~\cite{Devi:2013gya,Cirigliano:2017tqn}. In this section, we shall firstly start with a model-independent low-energy effective Lagrangian that contains all the potential NP operators contributing to the $\tau\to K_S\pi\nu_\tau$ decays, and analyze the tensor operator contribution to $A_{CP}^\tau$. Then, we shall discuss $A_{CP}^\tau$ in a scalar LQ scenario, which also contains the relevant operators.

\subsection{Model-independent analysis}

For the strangeness-changing hadronic $\tau$ decays, the most general model-independent effective Lagrangian at the characteristic scale $\mu_\tau=m_\tau$ can be written as~\cite{Cirigliano:2018dyk,Cirigliano:2017tqn,Rendon:2019awg}\footnote{This is adopted from the most general flavour-dependent low-energy effective Lagrangian governing the semi-leptonic $d^j\to u^i\ell \nu_\ell$ transitions, which can be found, for example, in Refs.~\cite{Antonelli:2008jg,Cirigliano:2009wk,Bhattacharya:2011qm}. It should be noted that, once the lepton-flavour universality is assumed, the effective couplings $\epsilon_{i}$ in Eq.~\eqref{eq:Efective_Lagrangian} would also receive the constraints from (semi-)leptonic kaon~\cite{Gonzalez-Alonso:2016etj} and hyperon~\cite{Chang:2014iba} decays.}
\begin{align}\label{eq:Efective_Lagrangian}
\mathcal{L}_{\rm eff}=&-\frac{G_{F}V_{us}}{\sqrt{2}}\,\Big\{(1+\epsilon_{L})\,\bar{\tau}\gamma_{\mu}
(1-\gamma_{5})\nu_{\tau}\cdot\bar{u}\gamma^{\mu}(1-\gamma_5)s+  \epsilon_{R}\,\bar{\tau}\gamma_{\mu}(1-\gamma_{5})\nu_{\tau}\cdot\bar{u}\gamma^{\mu}(1+\gamma_5)s\,\nonumber\\
&+\bar{\tau}(1-\gamma_{5})\nu_{\tau}\cdot\bar{u}\left[\epsilon_{S}-\epsilon_{P}\gamma_{5}\right]s+
\epsilon_{T}\,\bar{\tau}\sigma_{\mu\nu}(1-\gamma_{5})\nu_{\tau}\cdot\bar{u}\sigma^{\mu\nu}(1-\gamma_{5})s\Big\}
+\mathrm{h.c.}\,\nonumber\\[0.1cm]
=&-\frac{G_{F}V_{us}}{\sqrt{2}}\,(1+\epsilon_{L}+\epsilon_{R})\,\Big\{\bar{\tau}\gamma_{\mu}
(1-\gamma_{5})\nu_{\tau}\cdot\bar{u}\left[\gamma^{\mu}-(1-2\hat{\epsilon}_{R})\gamma^{\mu}\gamma_{5}\right]s\,\nonumber\\
&+\bar{\tau}(1-\gamma_{5})\nu_{\tau}\cdot\bar{u}\left[\hat{\epsilon}_{S}-\hat{\epsilon}_{P}\gamma_{5}\right]s+
2\hat{\epsilon}_{T}\bar{\tau}\sigma_{\mu\nu}(1-\gamma_{5})\nu_{\tau}\cdot\bar{u}\sigma^{\mu\nu}s\Big\}+
\mathrm{h.c.}\,,
\end{align}
where $G_{F}$ is the Fermi constant, and $\sigma^{\mu\nu}=\frac{i}{2}\,[\gamma^\mu,\gamma^\nu]$. The effective couplings $\epsilon_{i}$ parametrize the non-standard contributions and can be generally complex, with the SM case recovered when all $\epsilon_{i}=0$. We have also introduced the notations $\hat{\epsilon}_{i}=\epsilon_{i}/(1+\epsilon_L+\epsilon_R)$ for $i=R, S, P, T$, with the corresponding quark currents possessing definite parities and being therefore convenient to describe the vacuum to $K\pi$ matrix elements due to the parity conservation~\cite{Rendon:2019awg}. Here we have assumed Lorentz and $SU(3)_C\times U(1)_{\text{em}}$ invariance, as well as the absence of light non-standard particles when constructing $\mathcal{L}_{\rm eff}$\footnote{One should keep in mind that, unless some NP between $\mu_\tau=m_\tau$ and the electroweak scale $v=246~\text{GeV}$ is assumed, the low-energy effective Lagrangian given by Eq.~\eqref{eq:Efective_Lagrangian} comes generally from an $SU(2)$-invariant form~\cite{Buchmuller:1985jz,Grzadkowski:2010es,Alonso:2014csa}. This implies that the effective tensor operator contributing to $\tau\to K_S\pi\nu_\tau$ decays is also constrained by other processes, for example, by the neutron EDM and $D-\bar{D}$ mixing~\cite{Cirigliano:2017tqn}}. Right-handed and wrong-flavour neutrino contributions have also been neglected in Eq.~\eqref{eq:Efective_Lagrangian}, because they do not interfere with the SM amplitudes. 

Starting with Eq.~\eqref{eq:Efective_Lagrangian} and working in rest frame of the $\tau$ lepton, one can then obtain the differential decay width of the decay $\tau^-\to \bar{K}^0\pi^-\nu_\tau$~\cite{Cirigliano:2017tqn,Rendon:2019awg}
\begin{align}\label{eq:difw}
\frac{d \Gamma(\tau^-\to \bar{K}^0\pi^-\nu_\tau)}{d s}=&\frac{G_{F}^{2}\left|F_+(0)V_{u s}\right|^{2} m_{\tau}^{3} S_{\rm EW}}{384 \pi^{3} s}\,\left|1+\epsilon_{L}+\epsilon_{R}\right|^{2}\,\left(1-\frac{s}{m_{\tau}^{2}}\right)^{2} \lambda^{1 / 2}\left(s, M_{K}^{2}, M_{\pi}^{2}\right)\nonumber \\[0.2cm]
& \hspace{-1.8cm} \times \Big[X_{V A}+\Re e\hat{\epsilon}_{S} X_{S}+\Re e\hat{\epsilon}_{T} X_{\Re eT}+\Im m\hat{\epsilon}_{T} X_{\Im mT}+|\hat{\epsilon}_{S}|^2 X_{S^{2}}+|\hat{\epsilon}_{T}|^2 X_{T^{2}}\Big]\,,
\end{align}
where $s=(p_K+p_\pi)^2$, and $\lambda(s,M_K^2,M_\pi^2)=\left[s-(M_K+M_\pi)^2\right]\left[s-(M_K-M_\pi)^2\right]$. The product $\left|F_+(0)V_{us}\right|=0.21654(41)$ is determined most precisely from the analysis of semi-leptonic kaon decays~\cite{Antonelli:2010yf,Moulson:2017ive}. $S_{\rm EW}=1.0201(3)$ encodes the short-distance electroweak correction~\cite{Erler:2002mv}, and is simply written as an overall constant~\cite{Rendon:2019awg}. We have also introduced the following quantities:
\begin{align}
X_{V A}&=\frac{1}{2 s^{2}}\left[3|\tilde{F}_{0}(s)|^{2} \Delta_{K \pi}^{2}+|\tilde{F}_{+}(s)|^{2}\left(1+\frac{2 s}{m_{\tau}^{2}}\right) \lambda(s, M_{K}^{2}, M_{\pi}^{2})\right]\,,\label{eq:XVA}\\[0.2cm]
X_{S}&=\frac{3}{s\,m_{\tau}}|\tilde{F}_{0}(s)|^{2} \frac{\Delta_{K \pi}^{2}}{m_{s}-m_{u}}\,,\label{eq:XS}\\[0.2cm]
X_{\Re eT}&=-\frac{6}{s\,m_{\tau}}\left|\frac{F_T(0)}{F_+(0)}\right||\tilde{F}_{T}(s)|| \tilde{F}_{+}(s)|\,\cos\left[\delta_T(s)-\delta_+(s)\right] \lambda\left(s, M_{K}^{2}, M_{\pi}^{2}\right)\,,\label{eq:XRET}\\[0.2cm]
X_{\Im mT}&=-\frac{6}{s\,m_{\tau}}\left|\frac{F_T(0)}{F_+(0)}\right||\tilde{F}_{T}(s)|| \tilde{F}_{+}(s)|\,\sin\left[\delta_T(s)-\delta_+(s)\right] \lambda\left(s, M_{K}^{2}, M_{\pi}^{2}\right)\,,\label{eq:XIMT}\\[0.2cm]
X_{S^{2}}&=\frac{3}{2 m_{\tau}^{2}}|\tilde{F}_{0}(s)|^{2} \frac{\Delta_{K \pi}^{2}}{\left(m_{s}-m_{u}\right)^{2}}\,,\label{eq:XS2}\\[0.2cm]
X_{T^{2}}&=\frac{4}{s}\left|\frac{F_T(0)}{F_+(0)}\right|^2|\tilde{F}_{T}(s)|^{2}\left(1+\frac{s}{2 m_{\tau}^{2}}\right) \lambda\left(s,  M_{K}^{2},M_{\pi}^{2}\right)\,.\label{eq:XT2}
\end{align}
Here we have split $F_i(s)=F_i(0)\tilde F_i(s)$ (with $i=+,0,T$ corresponding to the vector, scalar, and tensor form factors, respectively) into $F_i(0)$ (form factors at the zero momentum transfer) and $\tilde{F}_i(s)$ (the corresponding normalized form factors), with $F_i(s)$ defined respectively as~\cite{Rendon:2019awg}
\begin{align}
\langle \bar{K}^0(p_K)\pi^-(p_\pi)|\bar s\gamma^\mu u|0\rangle&=\left[(p_K-p_\pi)^\mu-\frac{\Delta_{K\pi}}{s}q^\mu\right]F_+(s)+\frac{\Delta_{K\pi}}{s}q^\mu F_0(s)\,,\label{eq:vc}
\\[0.2cm]
\langle \bar{K}^0(p_K)\pi^-(p_\pi)|\bar s\,u|0\rangle&=\frac{\Delta_{K\pi}}{m_s-m_u}F_0(s)\,,\label{eq:sff}
\\[0.2cm]
\langle \bar{K}^0(p_K)\pi^-(p_\pi)|\bar s\sigma^{\mu\nu}u|0\rangle&=iF_T(s)\left(p_K^\mu p_\pi^\nu-p_K^\nu p_\pi^\mu\right)\,,\label{eq:tc}
\end{align}
where $q^\mu=(p_K+p_\pi)^\mu$, $\Delta_{K\pi}=M_K^2-M_\pi^2$, and the equation of motion has been used in Eq.~\eqref{eq:sff}.

The differential decay rate of the CP-conjugated process $\tau^+\to K^0\pi^+\bar{\nu}_\tau$ is obtained from Eq.~\eqref{eq:difw} by changing the sign of the term $\Im m\hat{\epsilon}_T$, which implies that only this term contributes to the direct CP asymmetry. From the definition of Eq.~\eqref{eq:ACPTau}, the direct CP asymmetry due to non-standard tensor interaction can be finally written as~\cite{Cirigliano:2017tqn,Rendon:2019awg}
\begin{align}
A_{CP}^{\tau}=\frac{2\,\Im m\hat{\epsilon}_{T}\, G_{F}^{2}| F_+(0)V_{u s}|^{2} S_{\rm EW}}{256\,\pi^{3}\, m_{\tau}^{2}\, \Gamma(\tau \rightarrow K_{S} \pi \nu_{\tau})}\left|\frac{F_T(0)}{F_+(0)}\right|\, &\int_{s_{K\pi }}^{m_{\tau}^{2}} d s\,\frac{\lambda^{3 / 2}\left(s, M_{K}^{2}, M_{\pi}^{2}\right)\left(m_{\tau}^{2}-s\right)^{2}}{s^{2}}\nonumber\\[0.2cm]
&\hspace{-2.5cm}\times |\tilde{F}_{+}(s)||\tilde{F}_{T}(s)|\,\sin \left[\delta_{T}(s)-\delta_{+}(s)\right]\,,
\end{align}
where $s_{K\pi }=(M_K+M_\pi)^2$, and $\delta_{T}(s)$ and $\delta_{+}(s)$ are the strong phases of the tensor and vector form-factors, respectively. The decay width $\Gamma(\tau^-\to K_S\pi^-\nu_\tau)$ as well as the branching ratio $\mathcal B(\tau^-\to K_S\pi^-\nu_\tau)=\Gamma(\tau^-\to K_S\pi^-\nu_\tau)/\Gamma_\tau$, with $\Gamma_\tau$ being the total decay width of the $\tau^-$ lepton, are obtained by integrating Eq.~\eqref{eq:difw} over $s$ from $s_{K\pi }$ to $m_\tau^2$.

\subsection{Analysis in the scalar LQ scenario}

In this subsection, we study the direct CP asymmetry in the scalar LQ scenario~\cite{Bauer:2015knc}, which was proposed by Bauer and Neubert to address the $R_{D^{(\ast)}}$, $R_K$, and $(g-2)_\mu$ anomalies, and can also generate the required tensor operator. In this scenario, only a single TeV-scale scalar LQ $\phi$, transforming as $(\mathbf{3}, \mathbf{1}, -\frac{1}{3})$ under the SM gauge group, is added to the SM particle content, and its couplings to fermions are described by the Lagrangian~\cite{Bauer:2015knc}
\begin{equation}
\mathcal L_{\rm int}^\phi=\bar Q_L^c{\boldsymbol\lambda}^Li\tau_2L\phi^{\ast}+\bar u_R^c{\boldsymbol\lambda}^R\ell_R\phi^{\ast}+{\rm h.c.}\,,\label{eq:Lint}
\end{equation}
where ${\boldsymbol\lambda}^{L,R}$ are the Yukawa coupling matrices in flavour space, $Q_L$ and $L$ denote the left-handed quark and lepton doublet, while $u_R$ and $\ell_R$ are the right-handed up-type quark and lepton singlet, with $\psi^c=C\bar{\psi}^T$ and $\bar{\psi}^c=\psi^T C$~($C=i\gamma^2\gamma^0$) being the charge-conjugated spinors.

With the SM as well as the tree-level $\phi$-mediated contributions included, the resulting effective Hamiltonian governing the $\tau \to K_S\pi\nu_\tau$ decays can be written as
\begin{align}
\mathcal H_{\text{eff}}\!=\!\frac{G_F V_{us}}{\sqrt2}\Big\{
&\left[1\!+\!C_V(\mu_\phi)\right]\,\bar\tau\gamma^\mu (1-\gamma_5)\nu_\tau\cdot\bar u\gamma_\mu (1-\gamma_5) s\!+\!C_S(\mu_\phi)\,\bar\tau(1-\gamma_5)\nu_\tau\cdot\bar u(1-\gamma_5) s\,\nonumber\\
&\!+\!C_T(\mu_\phi)\,\bar{\tau}\sigma^{\mu\nu}(1-\gamma_5)\nu_{\tau}\cdot\bar u\sigma_{\mu\nu}(1-\gamma_5) s\Big\}+
\mathrm{h.c.}\,,\label{eq:Hamiltonian}
\end{align}
where $C_V(\mu_\phi)$, $C_S(\mu_\phi)$ and $C_T(\mu_\phi)$ denote the Wilson coefficients of the corresponding operators at the matching scale $\mu_\phi=M_\phi$, and are given, respectively, as~\cite{Bauer:2015knc,Li:2016vvp}
\begin{align}
C_V(\mu_\phi)=&\frac{\lambda_{u\tau}^{L\ast}\lambda_{s\nu_{\tau}}^L}
{4\sqrt{2}G_FV_{us}M_{\phi}^2}\,,\nonumber\\
 C_{S}(\mu_\phi)=&-4\,C_{T}(\mu_\phi)=-\frac{\lambda_{u\tau}^{R\ast}\lambda_{s\nu_{\tau}}^L}
{4\sqrt{2}G_FV_{us}M_{\phi}^2}\,,\label{eq:WCs}
\end{align}
in which all the couplings $\lambda^{L,R}_{ij}$ are generally complex, with $i$ and $j$ denoting the flavours of quarks and leptons, respectively. In order to re-sum the potentially large logarithmic effects, these Wilson coefficients should be evolved down to the characteristic scale $\mu_\tau=m_\tau$. The vector current is conserved and hence the corresponding Wilson coefficient is scale independent, while the evolution of the scalar ($C_S$) and tensor ($C_T$) ones at the leading logarithmic approximation can be written schematically as~\cite{Dorsner:2016wpm}
\begin{align}
C_{S,T}(\mu_\tau)=R_{S,T}(\mu_\tau,\mu_\phi)\,C_{S,T}(\mu_\phi)\,,
\end{align}
with the corresponding evolution functions $R_{S,T}(\mu_\tau, \mu_\phi)$ given by
\begin{align}
R_{S,T}(\mu_\tau,\mu_\phi)\equiv\bigg[\frac{\alpha_s(m_b)}{\alpha_s(\mu_\tau)}\bigg]^{\frac{\gamma_{S,T}}{2\beta_0^{(4)}}}\,\bigg[\frac{\alpha_s(m_t)}{\alpha_s(m_b)}\bigg]^{\frac{\gamma_{S,T}}{2\beta_0^{(5)}}}\,\bigg[\frac{\alpha_s(\mu_{\phi})}{\alpha_s(m_t)}\bigg]^{\frac{\gamma_{S,T}}{2\beta_0^{(6)}}} \,,
\end{align}
where $\beta_0^{(n_f)}=11-2n_f/3$ is the leading-order coefficient of the QCD beta function, with $n_f$ being the number of active quark flavours, and $\gamma_S=-8$~\cite{Chetyrkin:1997dh} and $\gamma_T=8/3$~\cite{Gracey:2000am} are the leading-order anomalous dimensions of the scalar and tensor currents, respectively.

Matching the relevant terms of the effective Hamiltonian (Eq.~\eqref{eq:Hamiltonian}) onto that of the model-independent effective Lagrangian (Eq.~\eqref{eq:Efective_Lagrangian}) at the same scale $\mu_\tau=m_\tau$, we get
\begin{align}\label{eq:WCLQ_low}
\epsilon_L&=C_V(\mu_\tau),~\epsilon_R=0,~\epsilon_S=\epsilon_P=C_S(\mu_\tau),~\epsilon_T=C_T(\mu_\tau)\,,\nonumber\\[0.1cm]
\hat{\epsilon}_R&=0,~\hat{\epsilon}_T=\hat{C}_T\equiv\frac{C_T(\mu_\tau)}{1+C_V(\mu_\tau)}\,,
\nonumber\\[0.1cm]
\hat{\epsilon}_S&=\hat{\epsilon}_P=\hat{C}_S\equiv\frac{C_S(\mu_\tau)}{1+C_V(\mu_\tau)}=-4\,\frac{R_{S}
(\mu_\tau,\mu_\phi)}{R_{T}(\mu_\tau,\mu_\phi)}\,\hat{C}_T\,.
\end{align}
Due to the specific relation $C_{S}(\mu_\phi)=-4\,C_{T}(\mu_\phi)$ at the matching scale $\mu_\phi=M_\phi$, we are actually left with only one effective coupling $\hat C_T$ in the scalar LQ scenario. This feature makes a sensitive difference compared to the model-independent case, as will be discussed later.

\section{Form factors in \boldmath{$\tau\to K_S\pi\nu_\tau$} decays}
\label{sec:FFs}

To set bounds on the non-standard interactions, one needs to have a precise knowledge of the $K\pi$ form factors introduced in Eqs.~\eqref{eq:vc}--\eqref{eq:tc}. To this end, one of the most appropriate approaches is the dispersive representation of these form factors, which warrants the properties of unitarity and analyticity. In this section, we shall firstly recapitulate the $K\pi$ vector and scalar form factors, and then present a calculation of the tensor form factor in the context of $\chi$PT with tensor sources and R$\chi$T with both $K^\ast(892)$ and $K^\ast(1410)$ included.

\subsection{Brief review of the vector and scalar form factors}

For the normalized vector form factor $\tilde F_+(s)$, we shall adopt the optimal three-times subtracted dispersion relation~\cite{Boito:2008fq,Boito:2010me}
\begin{align}\label{eq:vff}
\tilde F_+(s)&=\text{exp}\left\lbrace \lambda_+^{\prime}\frac{s}{M_{\pi^-}^2}+\frac{1}{2}(\lambda_+^{\prime\prime}-\lambda_+^{\prime\,2})\frac{s^2}
{M_{\pi^-}^4}+\frac{s^3}{\pi}\int_{s_{K\pi}}^{s_{cut}} ds'\frac{\delta_+(s')}{(s')^3(s'-s-i\epsilon)}\right\rbrace\,,
\end{align}
where one subtraction constant is fixed by $F_+(0)=1$, while the other two,  $\lambda_+^{\prime}$ and $\lambda_+^{\prime\prime}$, describe respectively the slope and curvature of $\tilde F_+(s)$ when performing its Taylor expansion around $s=0$, and hence encode the dominant low-energy behaviour of $\tilde F_+(s)$. As the calculation of $\lambda_+^{\prime}$ and $\lambda_+^{\prime\prime}$ requires the perfect knowledge of the form-factor phase $\delta_{+}(s)$ up to infinity, which is unrealistic, it becomes more suitable to treat them as free parameters that capture our ignorance of the higher-energy part of the dispersion integral~\cite{Boito:2008fq,Boito:2010me}. The constants $\lambda_+^{\prime}$ and $\lambda_+^{\prime\prime}$ can then be determined by fitting to the experimental data~\cite{Boito:2008fq,Boito:2010me,Escribano:2014joa}\footnote{If the phase $\delta_{+}(s)$ were exactly known, these two constants can also be determined by the spectral sum rules dictated by the asymptotic behaviour of $\tilde F_+(s)$~\cite{Bernard:2011ae}, but in this case the three-times subtracted dispersion representation given by Eq.~\eqref{eq:vff} would reduce to the standard once-subtracted version.}. With such a procedure, the subtraction terms cannot cancel perfectly the polynomial terms coming from the dispersion integral, and the use of the three-times subtracted dispersion representation would thus spoil the asymptotic behaviour of the vector form factor in the limit $s\to\infty$~\cite{Brodsky:1973kr,Lepage:1979zb,Lepage:1980fj}. This deficiency is, however, considered to be acceptable, because the vector form factor is employed only up to about $\sqrt{s}\simeq 1.7~\text{GeV}$, which is still in the resonance region~\cite{Boito:2008fq,Boito:2010me}. The three-times subtracted dispersion representation of $\tilde F_+(s)$ has also been checked explicitly to be a decreasing function of $s$ within the entire range applied, which renders this approach credible~\cite{Boito:2010me}.  

The procedure proposed in Refs.~\cite{Boito:2008fq,Boito:2010me} is featured by the fact that the subtraction terms reduce the sensitivity of the dispersion integral to the higher-energy contribution, with the associated constants being less model dependent, and the impact of our ignorance of the form-factor phase $\delta_{+}(s)$ at relatively higher energies turns out to be very small. This makes it reasonable to determine the form-factor phase $\delta_{+}(s)$ in the context of R$\chi$T with the two vector resonances $K^\ast(892)$ and $K^\ast(1410)$ included~\cite{Boito:2008fq,Boito:2010me,Jamin:2006tk,Jamin:2008qg}. Notice that, in the elastic region below roughly $1.2~{\rm GeV}$, the phase $\delta_{+}(s)$ equals the scattering phase $\delta_1^{1/2}(s)$ of the $K\pi$ system with spin-$1$ and isospin-$1/2$, as required by Watson's theorem~\cite{Watson:1954uc}. The cut-off $s_{cut}$ introduced in Eq.~\eqref{eq:vff} is used to quantify the suppression of the higher-energy part of the integral, and the stability of the numerical results has been checked by varying $s_{cut}$ in the range $m_{\tau}<\sqrt{s_{cut}}<\infty$~\cite{Boito:2008fq,Boito:2010me}.

Detailed information on the $K\pi$ vector form factor can also be obtained from the measured $\tau\to K_S\pi\nu_\tau$ spectrum~\cite{Epifanov:2007rf}. This is, however, possible only for its modulus but not for its phase, as the extraction of the latter requires a fit function that preserves the analytic structure of the form factor. Indeed, the phase fitted via a superposition of Breit-Wigner functions with complex coefficients cannot be physical, as it does not vanish at threshold and violates Watson's theorem long before the resonance $K^\ast(1410)$ starts to play an effect~\cite{Cirigliano:2017tqn}. Thus, one cannot rely on the formalism developed in Ref.~\cite{Epifanov:2007rf} to study the CP asymmetry in $\tau\to K_S\pi\nu_\tau$ decays.

For the scalar form factor, a thorough description that takes into account analyticity, unitarity, the large-$N_C$ limit of QCD, as well as the couplings to $K\eta$ and $K\eta^\prime$ channels has been presented in Ref.~\cite{Jamin:2001zq} and later updated in Refs.~\cite{Jamin:2001zr,Jamin:2004re,Jamin:2006tj}. Here we shall employ such a coupled-channel dispersive representation, with the relevant numerical tables obtained via a combined analysis of the $\tau^-\to K_S\pi^-\nu_\tau$ and $\tau^-\to K^-\eta\nu_\tau$ decays~\cite{Escribano:2014joa}\footnote{We thank Pablo Roig for providing  us with these necessary numerical tables obtained in Ref.~\cite{Escribano:2014joa}.}.

\subsection{Calculation of the \boldmath{$K\pi$} tensor form factor}

Unlike for the vector and scalar form factors, there exists no experimental data to guide us to construct the tensor form factor, and we have to rely on theory to perform this task. In this subsection, following Refs.~\cite{Miranda:2018cpf,Rendon:2019awg}, we present a new calculation of the $K\pi$ tensor form factor.

\subsubsection{Result at the lowest chiral order of \boldmath{$\chi$}PT with tensor sources}

When the external tensor field $\bar{t}^{\mu\nu}=\sum\limits_{a=0}^{8}\frac{\lambda_a}{2}\bar{t}^{\mu\nu}_a$, with $\lambda_0=\sqrt{2/3}\,\mathbf{1}_{3\times3}$ and $\lambda_{1,\dots,8}$ being the eight Gell-Mann matrices, is switched on, the lowest-order ($\mathcal{O}(p^4)$ in the chiral counting) $\chi$PT Lagrangian can be written as~\cite{Cata:2007ns,Mateu:2007tr}
\begin{align}\label{eq:L4xpt}
\mathcal{L}_{4}^{\chi\text{PT}}=\Lambda_1\,\langle t_{+}^{\mu\nu}\,f_{+\mu\nu}\rangle -i\Lambda_2\,\langle t_{+}^{\mu\nu}\,u_{\mu}u_{\nu}\rangle + \Lambda_3\,\langle t_{+}^{\mu\nu}\,t^{+\mu\nu}\rangle +\Lambda_4\,{\langle t_{+}^{\mu\nu}\rangle}^2\,,
\end{align}
where $\langle\cdots\rangle$ denotes the trace in flavour space and, among the four operators, only the one with coefficient $\Lambda_2$ contributes to the $\tau\to K_S\pi\nu_\tau$ decays. The building blocks $t_{+}^{\mu\nu}=u^\dagger t^{\mu\nu} u^\dagger+u\,t^{\mu\nu\dagger} u$ and $u_{\mu}=i\left[u^\dagger(\partial_{\mu}-i r_{\mu})u- u(\partial_{\mu}-i l_{\mu})u^\dagger \right]$ are built out of the unitary non-linear representation of the pseudo Goldstone fields, $u(\phi^a)=\text{exp}\left(\frac{i}{2F_{\pi}} \phi^a\lambda^a\right)$~\cite{Coleman:1969sm,Callan:1969sn}, where $\phi^a=(\pi,K,\eta)$, $F_\pi=92.3(1)~{\rm MeV}$ is the physical pion decay constant~\cite{Tanabashi:2018oca}, and $l_{\mu}$ and $r_{\mu}$ are the left- and right-handed sources, respectively. The chiral tensor sources $t^{\mu\nu}$ and $t^{\mu\nu\dagger}$ are related to $\bar{t}^{\mu\nu}$ via~\cite{Cata:2007ns}
\begin{align}
\bar{t}^{\mu\nu}=P_{L}^{\mu\nu\alpha\beta}\,t_{\alpha\beta} + P_{R}^{\mu\nu\alpha\beta}\,t_{\alpha\beta}^{\dagger}\,, \qquad
t^{\mu\nu}=P_{L}^{\mu\nu\alpha\beta}\,\bar{t}_{\alpha\beta}\,,
\end{align}
in which $P_{R}^{\mu\nu\alpha\beta}=\frac{1}{4}\left(g^{\mu\alpha}g^{\nu\beta} -g^{\mu\beta}g^{\nu\alpha} +i \epsilon^{\mu\nu\alpha\beta}\right)$, with the convention $\epsilon^{0123}=+1$ for the Levi-Civit\'{a} tensor $\epsilon^{\mu\nu\alpha\beta}$, and the algebraic identity $\sigma^{\mu\nu}\gamma_5=\frac{i}{2}\epsilon^{\mu\nu\alpha\beta}\sigma_{\alpha\beta}$ has been used to get the relation $P_{L}^{\mu\nu\alpha\beta}=\left(P_{R}^{\mu\nu\alpha\beta}\right)^{\dagger}$.

Taking the functional derivative of Eq.~\eqref{eq:L4xpt} with respect to the tensor source $\bar{t}^{\mu\nu}$, with all the other external sources put to zero, expanding $u(\phi^a)$ in powers of $\phi^a$, and then taking the suitable hadronic matrix element, one can finally get~\cite{Garces:2017jpz,Miranda:2018cpf,Rendon:2019awg}
\begin{align}
 \left\langle \bar K^0(p_{K})\pi^-(p_{\pi})\left|\frac{\delta\mathcal L_4^{\chi\text{PT}}}{\delta\bar t_{\mu\nu}}\right|0\right\rangle=i\frac{\Lambda_2}{F_\pi^2}\left(p_{K}^\mu p_{\pi}^\nu-p_{K}^\nu p_{\pi}^\mu\right)\,.\label{eq:tensorff}
\end{align}
This, together with Eq.~\eqref{eq:tc}, fixes the normalization $F_T(0)=\frac{\Lambda_2}{F_\pi^2}$ at the lowest chiral order in $\chi$PT. Although the low-energy constant $\Lambda_2$ cannot be determined from the $\chi$PT itself, its value can be inferred either from other low-energy constants using the short-distance constraint Eq.~\eqref{eq:short} (see the next subsection for more details) or from the lattice result for the normalization $F_T(0)$~\cite{Baum:2011rm}. Here we shall resort to the lattice result, $F_T(0)=0.417(15)$~\cite{Baum:2011rm}, to determine $\Lambda_2=11.1\pm 0.4~\text{MeV}$, which is consistent within errors with that quoted in Refs.~\cite{Miranda:2018cpf,Hoferichter:2018zwu,Dekens:2018pbu,Tanabashi:2018oca} for the $\pi\pi$ channel\footnote{When comparing the values of $\Lambda_2$ quoted in Refs.~\cite{Miranda:2018cpf,Hoferichter:2018zwu,Dekens:2018pbu,Tanabashi:2018oca}, one should keep in mind the different conventions used for $\Lambda_2$. Our convention is consistent with that used in Refs.~\cite{Miranda:2018cpf,Cata:2007ns}.}. This value of $\Lambda_2$ will be used in our numerical analysis.

\subsubsection{Including the spin-1 resonances in the context of R$\chi$T}

As the invariant mass squared $s$ in $\tau\to K_S\pi\nu_\tau$ decays varies from the $K\pi$ threshold $s_{K\pi}$ up to $m_\tau^2$, contributions to the form factors from light resonances, giving therefore the $s$ dependence of these form factors, should also be included for a refined analysis. As the spin-$1$ resonances can be described equivalently by vector or anti-symmetric tensor fields~\cite{Ecker:1988te,Ecker:1989yg}, the same resonances that contribute to $F_+(s)$ will also appear in $F_T(s)$. To discuss the chiral couplings of these resonances to the pseudo Goldstone fields in the presence of tensor currents, we shall use the more conventional vector representation of these spin-$1$ degrees of freedom, named as model \textbf{II} prescription in Ref.~\cite{Ecker:1989yg}. Explicitly, the R$\chi$T Lagrangian that is linear in the octet vector field $\hat{V}_\mu$ and contains the couplings to the tensor source at the lowest chiral order can be constructed as~\cite{Ecker:1989yg}
\begin{align}\label{eq:LII}
\mathcal{L}_{\textbf{II}}=\mathcal{L}_{kin}(\hat{V}_\mu)-\frac{1}{2\sqrt{2}}\,\big(f_V\,\langle\hat{V}_{\mu\nu}f_{+}^{\mu\nu}\rangle+ig_V\,\langle\hat{V}_{\mu\nu}[u^\mu\,,u^\nu]\rangle\big)-f_V^T\,\langle \hat V_{\mu\nu}t_+^{\mu\nu}\rangle\,,
\end{align}
with the kinetic spin-$1$ part given by~\cite{Ecker:1989yg}
\begin{align}
\mathcal{L}_{kin}(\hat{V}_\mu) =-\frac{1}{4}\langle \hat{V}_{\mu\nu}\hat{V}^{\mu\nu}-2M_V^2\hat{V}_\mu\hat{V}^\mu\rangle\,,
\end{align}
where $\hat{V}_{\mu\nu}=\nabla_{\mu}\hat{V}_{\nu}-\nabla_{\nu}\hat{V}_{\mu}$, with the covariant derivative defined by $\nabla_{\mu}\hat{V}_{\nu}=\partial_\mu\hat V_\nu+[\Gamma_\mu,\hat V_\nu]$ and $\Gamma_\mu=\frac{1}{2}\left[u^\dagger\left(\partial_\mu-ir_\mu\right)u+u\left(\partial_\mu-il_\mu\right)u^\dagger\right]$. Here $f_{+}^{\mu\nu}=u\,F_L^{\mu\nu}\,u^\dagger+u^\dagger\,F_R^{\mu\nu}\,u$ is expressed in terms of the field strength tensors $F_{L}^{\mu \nu}=\partial^{\mu} l^{\nu}-\partial^{\nu} l^{\mu}-i\left[l^{\mu}, l^{\nu}\right]$ and $F_{R}^{\mu \nu}=\partial^{\mu} r^{\nu}-\partial^{\nu} r^{\mu}-i\left[r^{\mu}, r^{\nu}\right]$, which are associated with the non-abelian external fields $l_\mu$ and $r_\mu$, respectively. The last term in Eq.~\eqref{eq:LII} is added to describe the interactions between spin-$1$ vector resonances and external tensor fields. The three couplings $f_V$, $g_V$ and $f_V^T$ are all real and given, respectively, as~\cite{Ecker:1989yg}
\begin{equation}
 f_V=\frac{F_V}{M_V}=\frac{\sqrt{2}F_\pi}{M_V}\,,\quad g_V=\frac{G_V}{M_V}=\frac{F_\pi}{\sqrt{2}M_V}\,,\quad \langle 0|\bar u(0)\sigma_{\mu\nu}s(0)|V(p)\rangle=if_V^T(\epsilon_\mu p_\nu-\epsilon_\nu p_\mu)\,,
\end{equation}
where the first two result from the equivalence of models~\textbf{I} and \textbf{II} prescriptions for the spin-$1$ resonances~\cite{Ecker:1989yg}, while $f_V^T$ is determined from the one-particle to vacuum matrix element~\cite{Cata:2008zc}.

\begin{figure}[ht]
	\centering
	\includegraphics[width=0.55\textwidth]{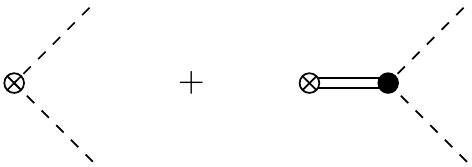}
	\caption{\label{fig:tensor interaction} \small Feynman diagrams contributing to $F_T(s)$ at the lowest chiral order of $\chi$PT (left) and from the lightest vector resonance ($K^{\ast}(892)$) exchange in the context of R$\chi$T (right). The crossed circles denote the insertion of a tensor current, and the blob represents the interaction vertex of a vector resonance (double line) with two pseudo-scalar mesons (dashed line). }
\end{figure}

With Eq.~\eqref{eq:LII} at hand, the effective action $S_{\textbf{II}}$ for a single vector meson exchange can then be written as~\cite{Ecker:1989yg}
\begin{equation}\label{eq:action}
S_{\textbf{II}}=\frac{1}{2}\int dx dy\langle J_{\textbf{II}}^{\mu\nu}(x)\Delta_{\mu\nu,\rho\sigma}^{\textbf{II}}(x-y)J_{\textbf{II}}^{\rho\sigma}(y)
\rangle\,,
\end{equation}
where the anti-symmetric current $J_{\textbf{II}}^{\mu\nu}$ and the resonance propagator $\Delta_{\mu\nu,\rho\sigma}^{\textbf{II}}$ are defined, respectively, by~\cite{Ecker:1989yg}
\begin{align}
J_{\textbf{II}}^{\mu\nu}&=\frac{1}{2\sqrt{2}}\,\left(f_V f_{+}^{\mu\nu} + ig_V \left[u^\mu,u^\nu\right]\right) + f_V^T\, t_+^{\mu\nu}\,,\\[0.2cm]
\Delta_{\mu\nu,\rho\sigma}^{\textbf{II}}(x-y)&=\int \frac{d^4k}{(2\pi)^4}\,\frac{e^{-ik\cdot(x-y)}}{k^2-M_V^2+i\epsilon}\,\left[g_{\mu\rho}\,k_\nu k_\sigma-g_{\mu\sigma}\,k_\nu k_\rho-(\mu\leftrightarrow\nu)\right]\,.
\end{align}
From the effective action $S_{\textbf{II}}$ given by Eq.~\eqref{eq:action}, one can easily derive the resonance contribution to the $K\pi$ tensor form factor due to the exchange of the lightest vector resonance $K^{\ast}(892)$~\cite{Miranda:2018cpf,Rendon:2019awg}. After including also the lowest-order $\chi$PT contribution given by Eq.~(\ref{eq:tensorff}), one then obtains the $K\pi$ tensor form factor
\begin{equation}\label{eq:FTs}
F_T(s)=\frac{\Lambda_2}{F_\pi^2}\left[1+\frac{\sqrt{2} f_V^Tg_V}{\Lambda_2}\frac{s}{M_{K^*}^2-s}\right]\,,
\end{equation}
with the corresponding Feynman diagrams in the context of $\chi$PT and R$\chi$T depicted in Fig.~\ref{fig:tensor interaction}.

In the large $N_C$ limit within $R\chi T$, although an infinite tower of resonances with the same quantum numbers as that of $K^{\ast}(892)$ should in principle be included in the computation of resonance-exchange amplitudes, it turns out that only the second state $K^{\ast}(1410)$ will actually play a crucial role in the inelastic region~\cite{Escribano:2013bca}. Accordingly, Eq.~\eqref{eq:FTs} should be changed to
\begin{equation}\label{eq:FTss}
F_T(s)=\frac{\Lambda_2}{F_\pi^2}\left[1+\frac{\sqrt{2}f_V^Tg_V}{\Lambda_2}\frac{s}{M_{K^*}^2-s}+\frac{\sqrt{2}f_V^{T\prime}g_V^{\prime}}{\Lambda_2}\frac{s}{M_{K^{*\prime}}^2-s}\right]\,,
\end{equation}
where $g_V^\prime$ and $f_V^{T\prime}$ are the counterparts of the corresponding unprimed couplings introduced in Eq.~\eqref{eq:LII}. We require further that $F_T(s)$ decreases at least as $1/s$ when $s\rightarrow\infty$~\cite{Brodsky:1973kr,Lepage:1979zb,Lepage:1980fj}, resulting in therefore the short-distance constraint
\begin{equation}\label{eq:short}
f_V^{T}g_V+f_V^{T\prime}g_V^\prime=\frac{\Lambda_2}{\sqrt{2}}\,.
\end{equation}
As the resonance exchange amplitudes are dominated by the first pole, one can then determine $\Lambda_2$ from Eq.~\eqref{eq:short} in terms of the known values of $f_{V}^T$ and $g_{V}$. Taking as inputs $F_\pi=92.3~\text{MeV}$ and $M_V=770~\text{MeV}$, we get $\Lambda_2\simeq\sqrt{2}f_V^Tg_V\simeq \sqrt{2}F_\pi^2/M_V \simeq 15.6~\text{MeV}$, which is compatible with the lattice result $\Lambda_2=11.1\pm 0.4~\text{MeV}$~\cite{Baum:2011rm}.

Analogous to the case for the vector form factor with the same two resonances included~\cite{Boito:2008fq}, the tensor form factor given by Eq.~(\ref{eq:FTss}) can also be rewritten as
\begin{align}\label{eq:FT3}
F_T(s)=\frac{\Lambda_2}{F_\pi^2}\left[\frac{M_{K^*}^2+\beta s}{M_{K^*}^2-s}-\frac{\beta s}{M_{K^{*\prime}}^2-s}\right]\,,
\end{align}
where the mixing parameter, $\beta=-\sqrt{2}f_V^{T\prime} g_V^\prime/\Lambda_2=\sqrt{2}f_V^{T}g_V/\Lambda_2-1$, is introduced to characterize the relative weight of the two resonances, and plays the same role as the parameter $\gamma$ does for the vector form factor~\cite{Boito:2008fq}. Although the parameter $\beta$ cannot be determined directly from data for the moment, we can estimate it from the fitted value of $\gamma$ with a three-times subtracted dispersion representation of the vector form factor~\cite{Boito:2008fq,Boito:2010me,Escribano:2014joa}. To this end, one needs firstly find out the relation between the R$\chi$T couplings $f_V^{T\prime}$ and $f_V^\prime\simeq\sqrt 2F_V^{\prime}$. The large-$N_C$ asymptotic analysis of $\langle VV\rangle$, $\langle TT\rangle$ and $\langle VT\rangle$ correlators suggests that a pattern with possible alternation in sign,
\begin{equation}\label{eq:xi}
 \xi_n=\frac{f_{Vn}^T}{f_{Vn}}=(-1)^n\frac{1}{\sqrt{2}}\,,
\end{equation}
exists for the whole $J^{PC}=1^{--}$ excited states~\cite{Cata:2008zc}. While $\xi_{K^*}$ is now confirmed to be positive~\cite{Becirevic:2003pn,Donnellan:2007xr,Allton:2008pn,Cata:2009dq,Dimopoulos:2011cf}, the sign of $\xi_{K^{*\prime}}$ can not be determined yet. Keeping both of these two possibilities, one can then derive the relation\footnote{Here we have used the relation  $g_V^\prime=G_V^\prime/M_{V^\prime}$, which also results from the equivalence of models \textbf{I} and \textbf{II} for the spin-$1$ degrees of freedom~\cite{Ecker:1989yg}.}
\begin{equation}
\frac{\beta}{\gamma}=\frac{-\sqrt{2}f_V^{T\prime} g_V^\prime/\Lambda_2}{-F_V^\prime G_V^\prime/F_\pi^2}=(-1)^n\frac{\sqrt{2}F_\pi^2}{M_{V^\prime}\Lambda_2 }\,,
\end{equation}
where $M_{V^\prime}\simeq M_{\rho^{\prime}}=1440~\text{MeV}$ in the limit of $SU(3)$ flavour symmetry. Thus, together with $\Lambda_2=11.1~\text{MeV}$ and $F_\pi=92.3~\text{MeV}$, one can express the parameter $\beta$ in terms of $\gamma$ via
\begin{equation}\label{eq:beta}
 \beta\simeq \pm 0.75\gamma\,,
\end{equation}
where both positive and negative signs of $\xi_{K^{*\prime}}$ will be considered in this paper. Intriguingly, our estimate, $\beta\simeq -0.75\gamma$, gives also a support for the assumption made in Ref.~\cite{Cirigliano:2017tqn} that the inelastic contributions to the phases of vector and tensor form factors are of similar size but potentially opposite in sign. 

As in the case for the vector form factor~\cite{Jamin:2006tk,Boito:2008fq}, the denominator in Eq.~\eqref{eq:FT3} should be modified by including the energy-dependent width $\gamma_{n}(s)$ (proportional to the imaginary part of the one-loop contribution in the context of $\chi$PT~\cite{Jamin:2006tk,Gasser:1984gg,Gasser:1984ux}) and also by shifting the pole mass (due to the real part of the loop contribution) of the resonances, as required by analyticity~\cite{Miranda:2018cpf}. After factoring out the normalization $F_T(0)$ at $s=0$, one arrives at the reduced tensor form factor $\tilde F_T(s)\equiv F_T(s)/ F_T(0)$, which is now given explicitly as
\begin{align}\label{eq:rTFF}
\tilde F_T(s)=\frac{m_{K^*}^2-\kappa_{K^*}\tilde{H}_{K\pi}(0)+\beta s}{D(m_{K^*},\gamma_{K^*})}-\frac{\beta s}{D(m_{K^{*\prime}},\gamma_{K^{*\prime}})}\,,
\end{align}
with the normalization $F_T(0)$ and the denominator $D(m_n,\gamma_n)$ defined, respectively, by
\begin{align}
F_T(0) &= \frac{\Lambda_2}{F_\pi^2}\frac{m_{K^*}^2}{m_{K^*}^2-\kappa_{K^*}\tilde{H}_{K\pi}(0)}\,,\\[0.2cm]
D(m_n,\gamma_n)&\equiv m_{n}^2-s-\kappa_n \Re e\tilde H_{K\pi}(s)-im_{n}\gamma_{n}(s)\,.
\end{align}
Here the parameters $m_{n}$ and $\gamma_{n}$ denote respectively the unphysical mass and width, to be distinguished from the physical mass $M_{n}$ and width $\Gamma_{n}$ that are obtained from the pole position in the complex $s$-plane~\cite{Jamin:2006tk,Boito:2008fq}. Explicit expressions for the one-loop function $\tilde{H}_{K\pi}(s)$, the energy-dependent width $\gamma_n(s)$, and the dimensional constant $\kappa_{n}$ can be found in Refs.~\cite{Jamin:2006tk,Boito:2008fq}.

Our result of $\tilde F_T(s)$ given by Eq.~\eqref{eq:rTFF} is quite similar to that of the reduced vector form factor $\tilde F_+(s)$ obtained in Ref.~\cite{Boito:2008fq}, except for the different normalization factors as well as the different relative weight parameters introduced to characterize the inelastic contributions. In the elastic region below roughly $1.2~{\rm GeV}$~\cite{Boito:2008fq}, one needs to set $\beta=\gamma=0$ and hence obtains $\tilde F_T(s)=\tilde F_+(s)$, which then implies that $\delta_T(s)=\delta_+(s)$, as required by the unitarity relation and the fact that the $K^\ast(892)$ resonance is described equivalently by a vector or an anti-symmetric tensor field~\cite{Ecker:1989yg,Ecker:1988te}. Furthermore, according to Watson's theorem~\cite{Watson:1954uc}, the phases of both $F_+(s)$ and $F_T(s)$ in the elastic region should coincide with the $P$-wave $K\pi$ phase shift $\delta_1^{1/2}(s)$. In such a case, no direct CP asymmetry in $\tau\to K_S\pi\nu_\tau$ decays will be predicted due to the lack of strong phase difference between vector and tensor form factors~\cite{Cirigliano:2017tqn}. Beyond the elastic region, however, a non-zero strong phase difference can be generated due to the different relative weight parameters in these two form factors, as will be shown in the next subsection.

\subsubsection{Dispersive representation of the tensor form factor}

In order to connect all the information on the form factors inferred from $\chi$PT at low energies, from the resonance dynamics in the intermediate energy region ($\mathcal{O}(1~\text{GeV})$), as well as from the short-distance QCD properties in the asymptotic regime~\cite{Brodsky:1973kr,Lepage:1979zb,Lepage:1980fj}, one can resort to the dispersive representation of the form factors, which fulfills the analyticity and unitarity requirements~\cite{Guerrero:1997ku,Pich:2001pj,Dumm:2013zh} and, at the same time, suppresses  the less-known higher-energy contributions~\cite{Boito:2008fq,Boito:2010me}.

In the elastic region below roughly $1.2~{\rm GeV}$, the dispersion relation for the vector and tensor form factors admits the well-known Omn\`es solution~\cite{Cirigliano:2017tqn}
\begin{align}\label{eq:unitarity}
F_{+,ela}(s)=F_+(0)\,\Omega(s)\,,\qquad F_{T,ela}(s)=F_T(0)\,\Omega(s)\,,
\end{align}
with the Omn\`es factor~\cite{Omnes:1958hv} given by
\begin{align}\label{eq:omnes}
\Omega(s)=\text{exp}\left[\frac{s}{\pi}\int_{s_{K\pi}}^\infty ds'\frac{\delta_1^{1/2}(s')}{s'(s'-s-i\epsilon)}\right]\,,
\end{align}
where the relation $\delta_T(s)=\delta_+(s)=\delta_1^{1/2}(s)$ in the elastic region has been used. As Watson's final-state interaction theorem is no longer valid starting from the threshold of inelastic states (most notably $K\pi\pi$~\cite{Cirigliano:2017tqn}), one must find a sensible way to determine the strong phase difference in the inelastic region, so as to predict a non-zero direct CP asymmetry in $\tau\to K_S\pi\nu_\tau$ decays. In this regard, our expression of $\tilde F_T(s)$ given by Eq.~\eqref{eq:rTFF} and that of $\tilde F_+(s)$ given by Eq.~(4.1) in Ref.~\cite{Boito:2008fq} are advantageous, because they remain valid even beyond the elastic approximation and the two form-factor phases can be calculated from the relations~\cite{Boito:2008fq,Jamin:2006tk,Escribano:2014joa}
\begin{align}\label{eq:ffphase}
\tan\delta_{T}(s)=\frac{\Im m[\tilde F_{T}(s)]}{\Re e[\tilde F_{T}(s)]}\,,\qquad
\tan\delta_{+}(s)=\frac{\Im m[\tilde F_{+}(s)]}{\Re e[\tilde F_{+}(s)]}\,,
\end{align}
in which the inelastic effects are indicated by the mixing parameters $\beta$ and $\gamma$, respectively.

It should be noted that the form-factor phases given by Eq.~\eqref{eq:ffphase} are valid only in the $\tau$-decay region $s_{K\pi}<s<m_{\tau}^2$. For the higher-energy region, these phases become generally unknown, but should be guided smoothly to $\pi$ (modulo $2\pi$) according to the asymptotic behaviour of the form factors at large $s$~\cite{Brodsky:1973kr,Lepage:1979zb,Lepage:1980fj}. Our ignorance of the form-factor phases at relatively higher energies also makes the numerical implementation of the dispersive integrals sensitive to the choice of the cut-off $s_{cut}$. For the vector form factor, once the three-times subtracted dispersion representation (see Eqs.~\eqref{eq:vff}) is adopted, the impact of this deficiency would be marginal, implying that the higher-energy contribution is well suppressed~\cite{Boito:2008fq,Boito:2010me,Escribano:2014joa}. For example, an input with a larger error band, $\delta_{+}(s)=\pi\pm\pi$, at $s\geq s_{cut}$ has been assumed in Ref.~\cite{Bernard:2011ae}, but the use of a three-times subtracted dispersion relation makes the integrand converge rapidly, and hence the result becomes almost insensitive to this large error assignment. Furthermore, the stability of the fit results has been checked explicitly by varying the cut-off $s_{cut}$ in a wide range $m_{\tau}<\sqrt{s_{cut}}<\infty$~\cite{Boito:2008fq,Boito:2010me}. It has  also been demonstrated that the choice $s_{cut}=4~\text{GeV}^2$ is preferred, because such a cut-off is, on the one hand, large enough to not spoil the \textit{priori} infinite interval of the dispersive integral and to avoid the spurious singularity effect generated at $s=s_{cut}$ and, on the other hand, low enough to give a good description of the form-factor phase within the interval considered~\cite{Gonzalez-Solis:2019iod}. Due to the lack of precise low-energy information on the tensor interaction, however, one cannot apply these strategies to the tensor form factor~\cite{Miranda:2018cpf,Rendon:2019awg}. Consequently, we shall simply use the once-subtracted dispersive representation~\cite{Cirigliano:2017tqn,Miranda:2018cpf,Rendon:2019awg}
\begin{align}\label{eq:tff}
\tilde F_T(s)&=\text{exp}\left\lbrace  \frac{s}{\pi}\int_{s_{K\pi}}^\infty ds'\frac{\delta_T(s')}{s'(s'-s-i\epsilon)}\right\rbrace\,,
\end{align}
together with the following simple model for the phase $\delta_{T}(s)$~\cite{Bernard:2009zm}:
\begin{align}\label{eq:nT}
\delta_T(s)=\begin{cases}\arctan{\left[\frac{\Im m\tilde F_T(s)}{\Re e\tilde F_T(s)}\right]}\,, & s_{K\pi}<s<s_{cut} \\[0.2cm]
n_T\pi\,, & s\geq s_{cut}\end{cases}\,,
\end{align}
where the phase is now made explicit even in the inelastic region, instead of the assumed relation $\delta_T(s)=-\delta_+(s)$ in the same region~\cite{Cirigliano:2017tqn}. We have also introduced the quantity $n_T$, with its deviation from unit, to account for our estimate of the uncertainty resulting from the  higher-energy contributions. In addition, the default choice with $s_{cut}=4~\text{GeV}^2$ and $\delta_T(s)=\pi$ for $s>s_{cut}$ will be assumed in our numerical analysis.

\begin{figure}[t]
	\centering
	\includegraphics[width=3.15in]{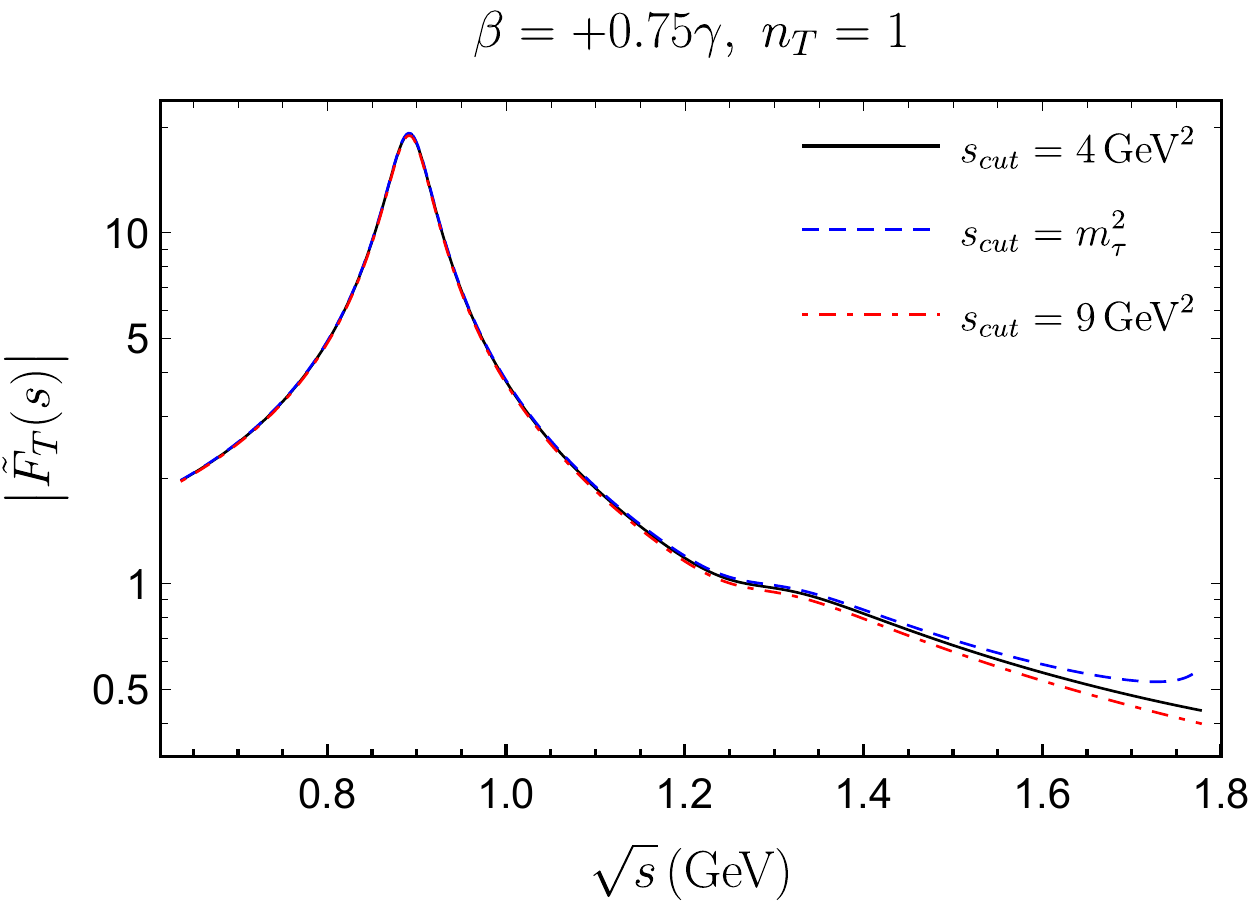}
	\hspace{0.15in}
	\includegraphics[width=3.15in]{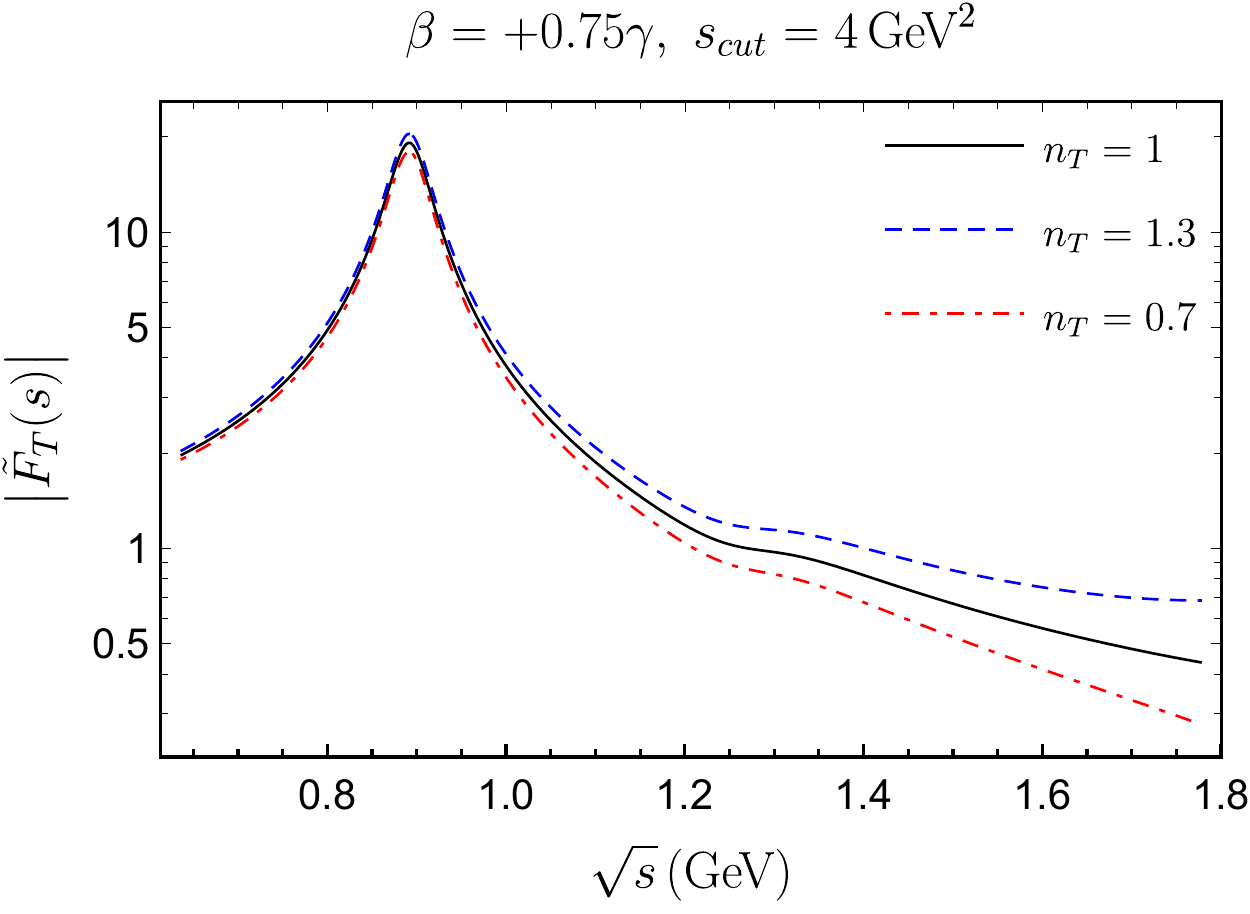}
	\caption{\label{fig:moduluserror} \small Dependence of the modulus of the normalized tensor form factor on $s_{cut}$ with fixed $n_T=1$ (left) and on $n_T$ with fixed $s_{cut}=4~\text{GeV}^2$, in the $\beta=+0.75\gamma$ case.}
\end{figure}

To estimate the systematic uncertainty associated with our model for the tensor form factor, we proceed as follows\footnote{These results are shown only for the purpose of making a comparison with that given in Ref.~\cite{Rendon:2019awg}, and will not be considered in the subsequent numerical analysis.}: by fixing $n_T=1$ to see the sensitivity of the modulus of the normalized tensor form factor with respect to $s_{cut}$, with the three choices $s_{cut}=m_\tau^2$, $4$ and $9~\text{GeV}^2$, and by fixing $s_{cut}=4~\text{GeV}^2$ to see the sensitivity with respect to $n_T$, with the three choices $n_T=1$, $1.3$ and $0.7$. Our numerical results with $\beta=+0.75\gamma$ (the case with $\beta=-0.75\gamma$ is quite similar) are shown in Fig.~\ref{fig:moduluserror}, from which it can be seen that, in our model, the modulus of the normalized tensor form factor is almost insensitive to the choice of the cutoff $s_{cut}$ when fixing $n_T=1$, while it becomes rather sensitive to the choice of $n_T$ when fixing $s_{cut}=4~\text{GeV}^2$, especially in the higher-energy region. This implies that the once-subtracted dispersive representation is not optimal, as is generally expected. But the lack of data sensitive to the tensor form factor makes it impossible to increase the number of subtractions for the moment.

\begin{figure}[ht]
	\centering
	\includegraphics[width=3.15in]{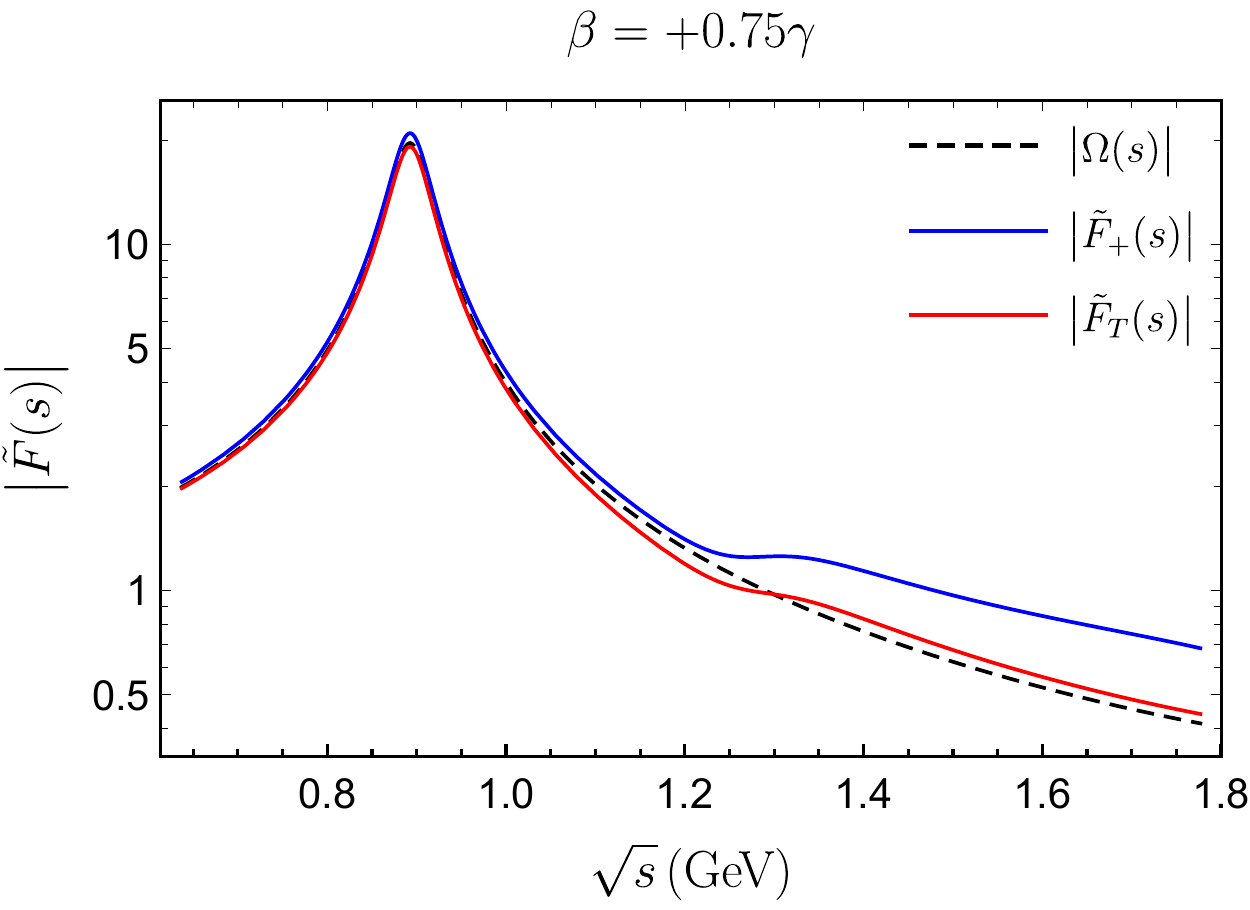}
	\hspace{0.15in}
	\includegraphics[width=3.15in]{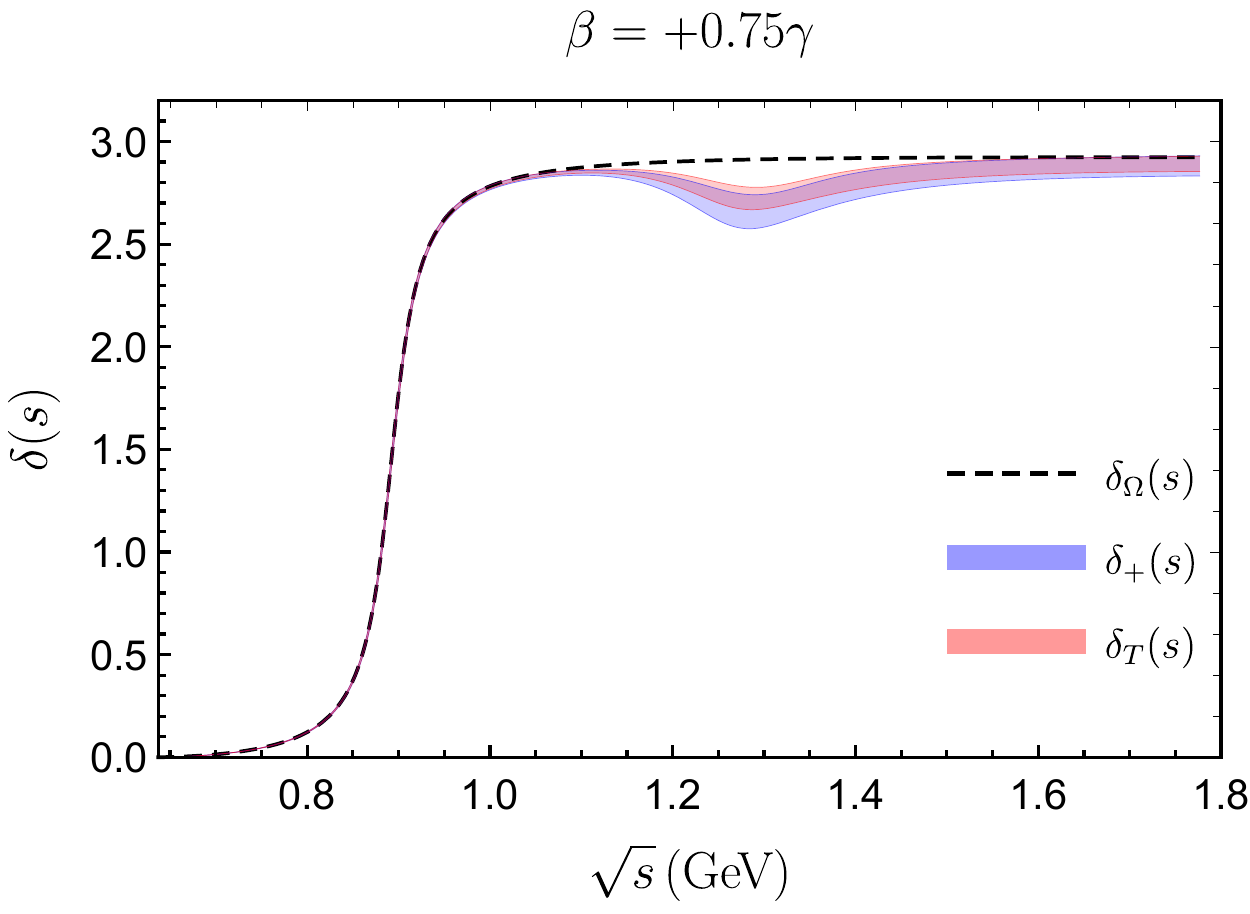}
	\caption{\label{fig:positive} \small Energy dependence of the moduli (left) and phases (right) of the normalized form factors, compared with the ones predicted by the Omn\`es factor $\Omega(s)$, with the bands resulting from the uncertainties of the input parameters, in the $\beta=+0.75\gamma$ case.}
\end{figure}

\begin{figure}[t]
	\centering
	\includegraphics[width=3.15in]{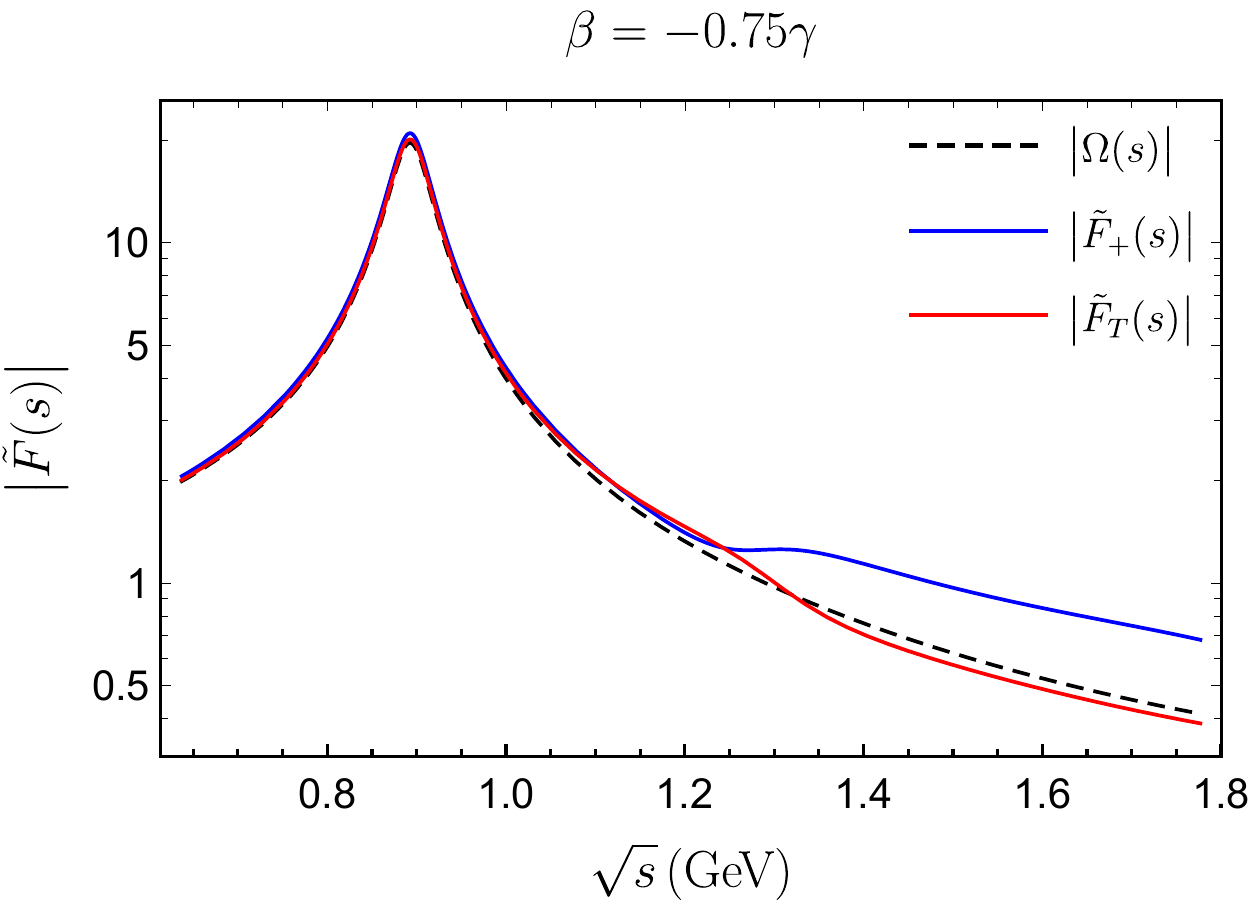}
	\hspace{0.15in}
	\includegraphics[width=3.15in]{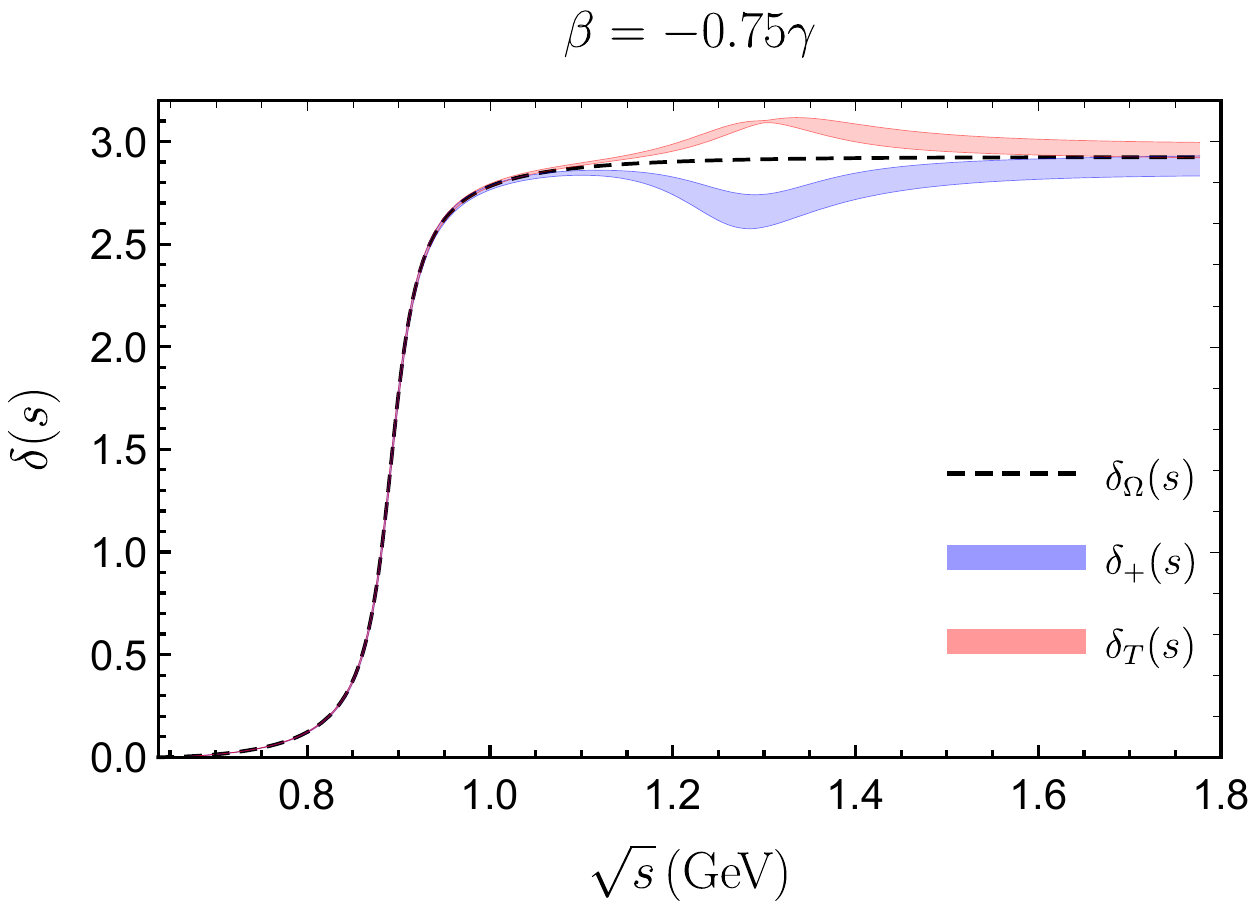}
	\caption{\label{fig:negative}\small Same as in Fig.~\ref{fig:positive} but in the $\beta=-0.75\gamma$ case.}
\end{figure}

To see the behaviours of the vector and tensor form factors both in the elastic and in the inelastic region, we show their moduli and phases as well as the ones predicted by the Omn\`es factor $\Omega(s)$ in Figs.~\ref{fig:positive} and \ref{fig:negative}, corresponding respectively to the two different choices given by Eq.~\eqref{eq:beta}. As the cut-off $s_{cut}$ has been fixed at $s_{cut}=4~\text{GeV}^2$, we consider the uncertainties of the form-factor phases only from the input parameters. From Figs.~\ref{fig:positive} and \ref{fig:negative}, one can see that both the moduli and the phases of the normalized form factors are consistent with the ones obtained from $\Omega(s)$ in the energy region up to about $1.2~\text{GeV}$, which is roughly the threshold of the inelastic region. The deviations from the ones predicted by $\Omega(s)$ in the higher-energy region, on the other hand, serve as an indication of the size of the inelastic contribution from the second resonance~\cite{Cirigliano:2017tqn}. It is also observed that, unlike in the case of the vector form factor, the modulus of the tensor form factor is almost unaffected by the inelastic effect, and is therefore similar to that obtained with $\Omega(s)$. The inelastic effects on the form-factor phases are, however, rather significant, and a strong phase difference in the inelastic region is indeed obtained, especially in the $\beta=-0.75\gamma$ case. This is welcome for resolving the CP anomaly observed in $\tau\to K_S\pi\nu_\tau$ decays, as will be discussed in the next section.

\section{Numerical results and discussions}
\label{sec:numerical analysis}

In this section, we discuss the numerical effects of the two NP scenarios introduced in section~\ref{sec:new physics} on the branching ratio $\mathcal{B}(\tau^-\to K_S\pi^-\nu_\tau)$ and the CP asymmetry $A_{CP}(\tau\to K_S\pi\nu_\tau)$. For each observable, the theoretical uncertainties are obtained by varying each input parameter within the corresponding range and then adding the individual errors in quadrature~\cite{Hocker:2001xe,Charles:2004jd,Li:2013vlx,Jung:2012vu}.

\subsection{Results in the model-independent framework}

As the $\tau^-\to \bar{K}^0\pi^-\nu_\tau$ decay width, which is obtained by integrating Eq.~\eqref{eq:difw} over the invariant mass squared $s$ within the kinematic regime $s_{K\pi}\leq s \leq m_{\tau}^2$, depends on the non-standard scalar and tensor interactions, it could also set bounds on these effective couplings~\cite{Devi:2013gya,Dhargyal:2016kwp,Dhargyal:2016jgo,Cirigliano:2017tqn,Rendon:2019awg,Dighe:2019odu}. In order to enhance the sensitivity to these non-standard interactions, we introduce the observable~\cite{Rendon:2019awg},
\begin{align}\label{eq:Delta}
\Delta\equiv\frac{\Gamma-\Gamma_0}{\Gamma_0}=a\,\Re e\hat{\epsilon}_S+b\,\Re e\hat{\epsilon}_T+c\,\Im m\hat{\epsilon}_T+d|\hat{\epsilon}_S|^2+e|\hat{\epsilon}_T|^2\,,
\end{align}
which is defined as the relative shift induced by these interactions, with $\Gamma$ and $\Gamma_0$ standing for the $\tau\to K_S\pi\nu_\tau$ decay widths with and without these non-standard contributions, respectively. The coefficients $a$, $b$, $c$, $d$ and $e$ are calculated, respectively, to be
\begin{align}\label{eq:aecoeff}
&a\in[0.27,0.34]\,,~\quad\quad d\in[0.84,1.12]\,,\\
&b\in[-4.46,-4.02]\,,\quad c\in[-0.005,0.015]\,,\quad e\in[6.0,7.4]\,,\quad \mbox{for $\beta=+0.75\gamma$}\,,\\
&b\in[-4.68,-4.24]\,,\quad c\in[0.026,0.046]\,,\qquad e\in[6.8,8.3]\,,\quad \mbox{for $\beta=-0.75\gamma$}\,.
\end{align}
It can be seen that our values of the coefficients $a$ and $d$, characterizing respectively the linear and the quadratic term of the non-standard scalar contributions, are consistent with that of $\alpha$ and $\gamma$ obtained in Ref.~\cite{Rendon:2019awg}, while the values of the  tensor coefficients are quite different due to the different forms of the $K\pi$ tensor form factor used. The numerical difference between scalar ($a$ and $d$) and tensor ($b$ and $e$) coefficients by about one order of magnitude implies a slightly larger sensitivity to the tensor than to the scalar contribution, as noted already in Ref.~\cite{Rendon:2019awg}. It is also observed that, although the real part of the interference between vector and tensor contributions is of similar magnitude as the pure tensor term, the imaginary part of the interference is almost negligible for both the $\beta=+0.75\gamma$ and $\beta=-0.75\gamma$ cases. This can be understood from the fact that the real and the imaginary part of this interference term are proportional to $\Re e[F_T(s)F_+(s)^{\ast}]$ and $\Im m[F_T(s)F_+(s)^{\ast}]$ (see Eqs.~\eqref{eq:XRET} and \eqref{eq:XIMT}), which in the elastic region are reduced to $\sim|F_T(s)||F_+(s)|$ and $\sim0$, respectively, with $|F_T(s)||F_+(s)|$ being of similar size as $|F_T(s)|^2$~\cite{Rendon:2019awg}. However, since only the imaginary part contributes to the direct CP asymmetry, its non-zero value is crucial in determining the observable $A_{CP}(\tau\to K_S\pi\nu_\tau)$.

For the CP asymmetry $A_{CP}(\tau\to K_S\pi\nu_\tau)$, the following subtle points should be clarified~\cite{BABAR:2011aa,Dighe:2019odu}. As the signal channel $\tau^-\to\pi^- K_S(\geq0\pi^0)\nu_\tau$ (C1) is contaminated by the two background channels $\tau^-\to K^- K_S(\geq0\pi^0)\nu_\tau$ (C2) and $\tau^-\to\pi^- K^0\bar K^0\nu_\tau$ (C3), the measured decay-rate asymmetry, $\mathcal{A}=(-0.27\pm0.18\pm0.08)\%$, by the BaBar collaboration~\cite{BABAR:2011aa} is actually related to the signal asymmetry $A_1$ as well as the two background asymmetries $A_2$ and $A_3$ via~\cite{BABAR:2011aa}
\begin{align}\label{eq:ACPextract}
\mathcal{A}&=\frac{f_1\,A_1+f_2\,A_2+f_3\,A_3}{f_1+f_2+f_3}\,\nonumber\\[0.2cm]
&=\frac{f_1-f_2}{f_1+f_2+f_3}\,A_{Q}\,,
\end{align}
where $f_1$, $f_2$ and $f_3$ denote the fractions of the channels C1, C2 and C3 in the total selected sample, with the corresponding numbers given in Table I of Ref.~\cite{BABAR:2011aa}. Within the SM, $A_1=-A_2$ because the $K_S$ state is produced via a $\bar{K}^0$ in channel C1 but via a $K^0$ in channel C2, and $A_3=0$ because of the cancellation between the CP asymmetries due to the $K^0$ and $\bar{K}^0$ states in channel C3. To extract the CP asymmetry $A_{Q}$ given by Eq.~\eqref{eq:ACP_Exp}, from the measured decay-rate asymmetry $\mathcal{A}$, these relations between $A_1$, $A_2$ and $A_3$ have been assumed by the BaBar collaboration~\cite{BABAR:2011aa}, as given by the second line in Eq.~\eqref{eq:ACPextract}. In the presence of NP contributions, however, $A_1\neq -A_2$ in general, and any theoretical prediction should be, therefore, compared with the measured quantity $\mathcal A$, instead of $A_Q$~\cite{Dighe:2019odu}. Assuming the NP contribution affects only the channel C1, we can then write the three CP asymmetries as~\cite{Dighe:2019odu}
\begin{align}
A_1&=A_1^{\text{SM}}+A_1^{\text{NP}}=A_{CP}^{\text{SM}}+A_1^{\text{NP}}\,,\nonumber\\[0.2cm]
A_2&=A_2^{\text{SM}}=-A_1^{\text{SM}}=-A_{CP}^{\text{SM}}\,,\nonumber\\[0.2cm]
A_3&=A_3^{\text{SM}}=0\,,
\end{align}
where the SM prediction $A_{CP}^{\text{SM}}=(0.36\pm0.01)\%$ is obtained after taking into account the $K_S\to \pi^+\pi^-$ decay-time dependence of the event selection efficiency~\cite{BABAR:2011aa}. Combing $A_{CP}^{\text{SM}}$ with the measured decay-rate asymmetry $\mathcal{A}=(-0.27\pm0.18\pm0.08)\%$~\cite{BABAR:2011aa}, we can therefore obtain constraint on $A_1^{\text{NP}}$, and then on the non-standard tensor coupling.

\begin{figure}[t]
  \centering
  \includegraphics[width=2.16in]{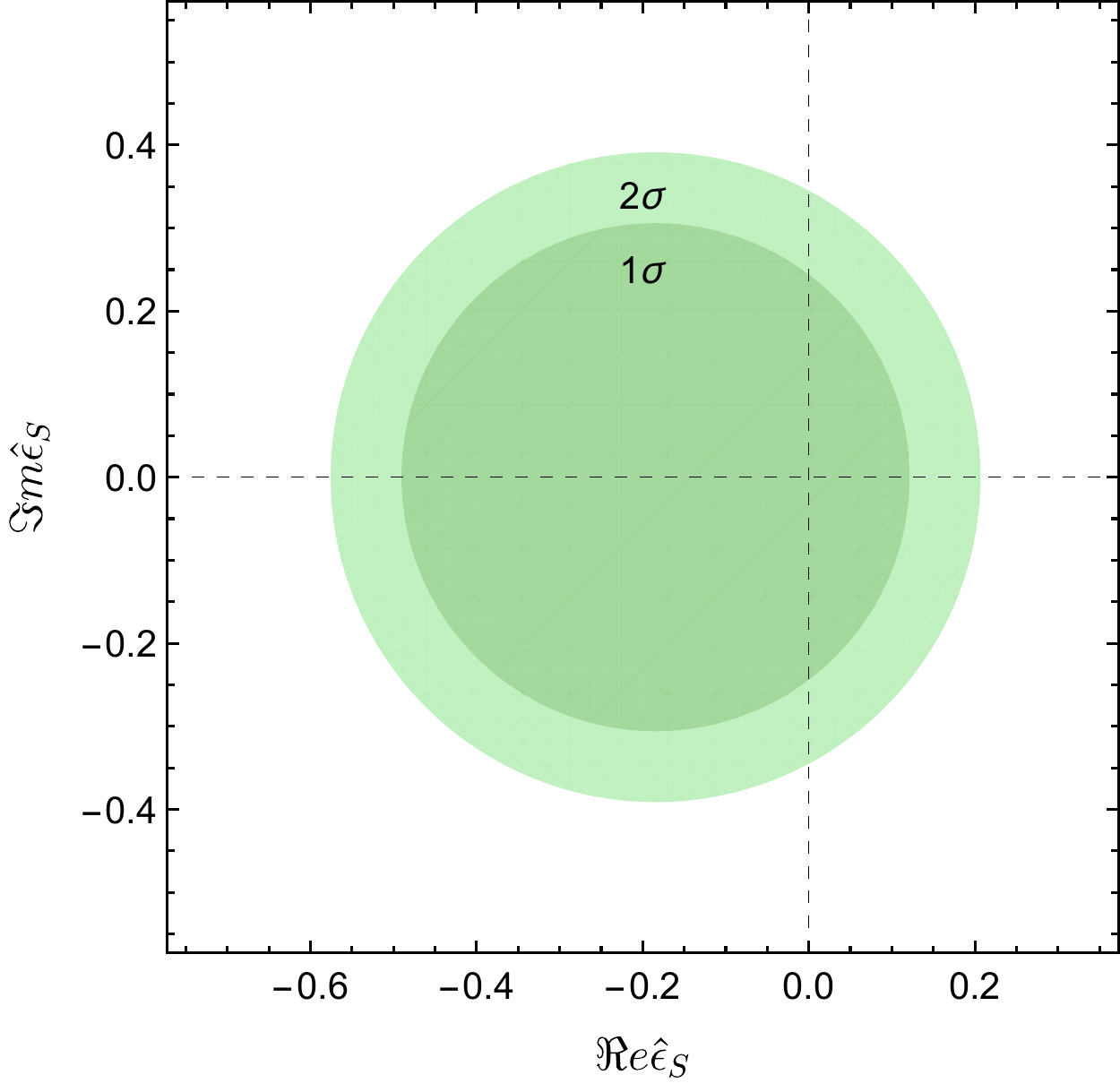}~
  \includegraphics[width=2.16in]{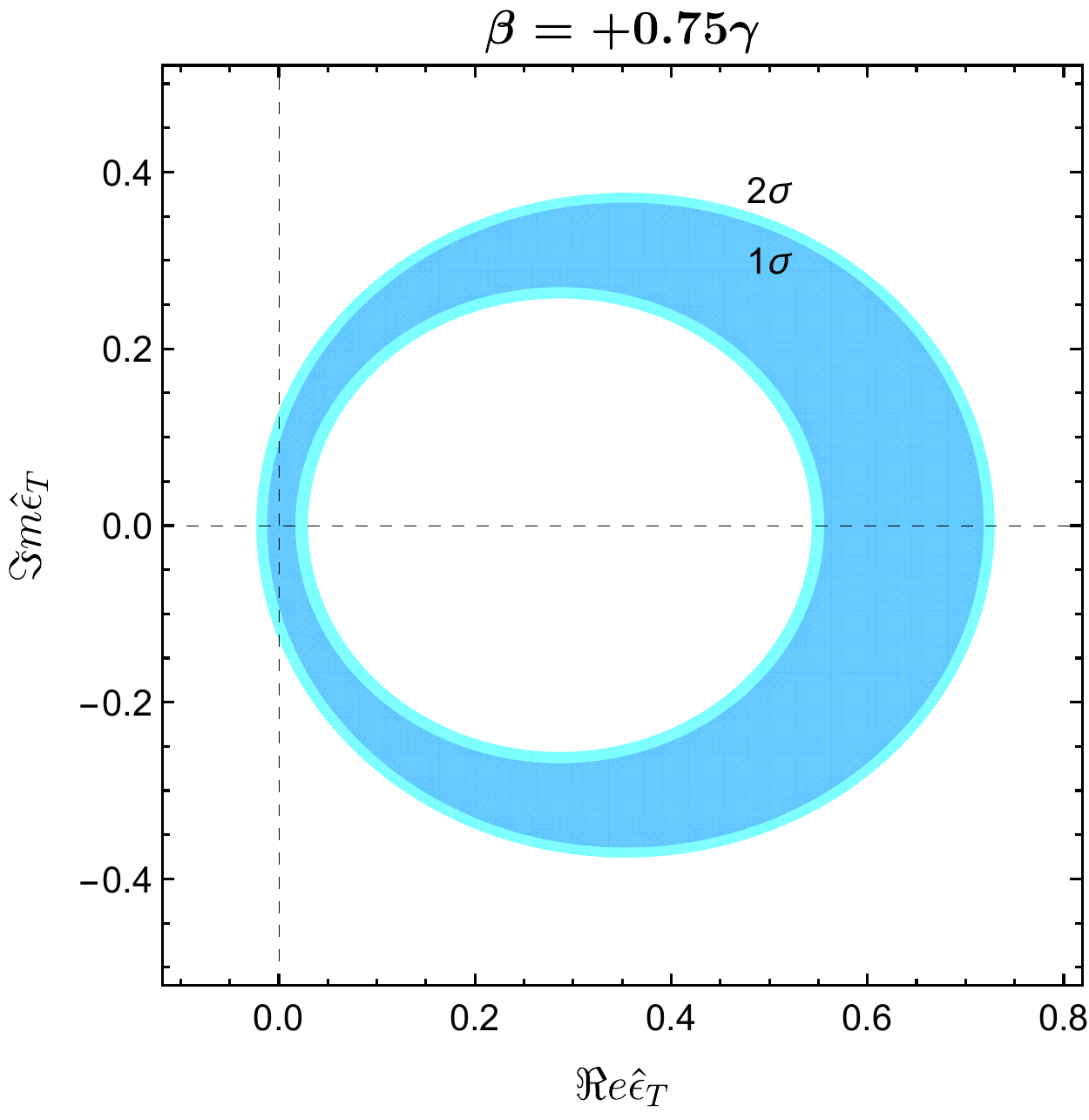}~
  \includegraphics[width=2.16in]{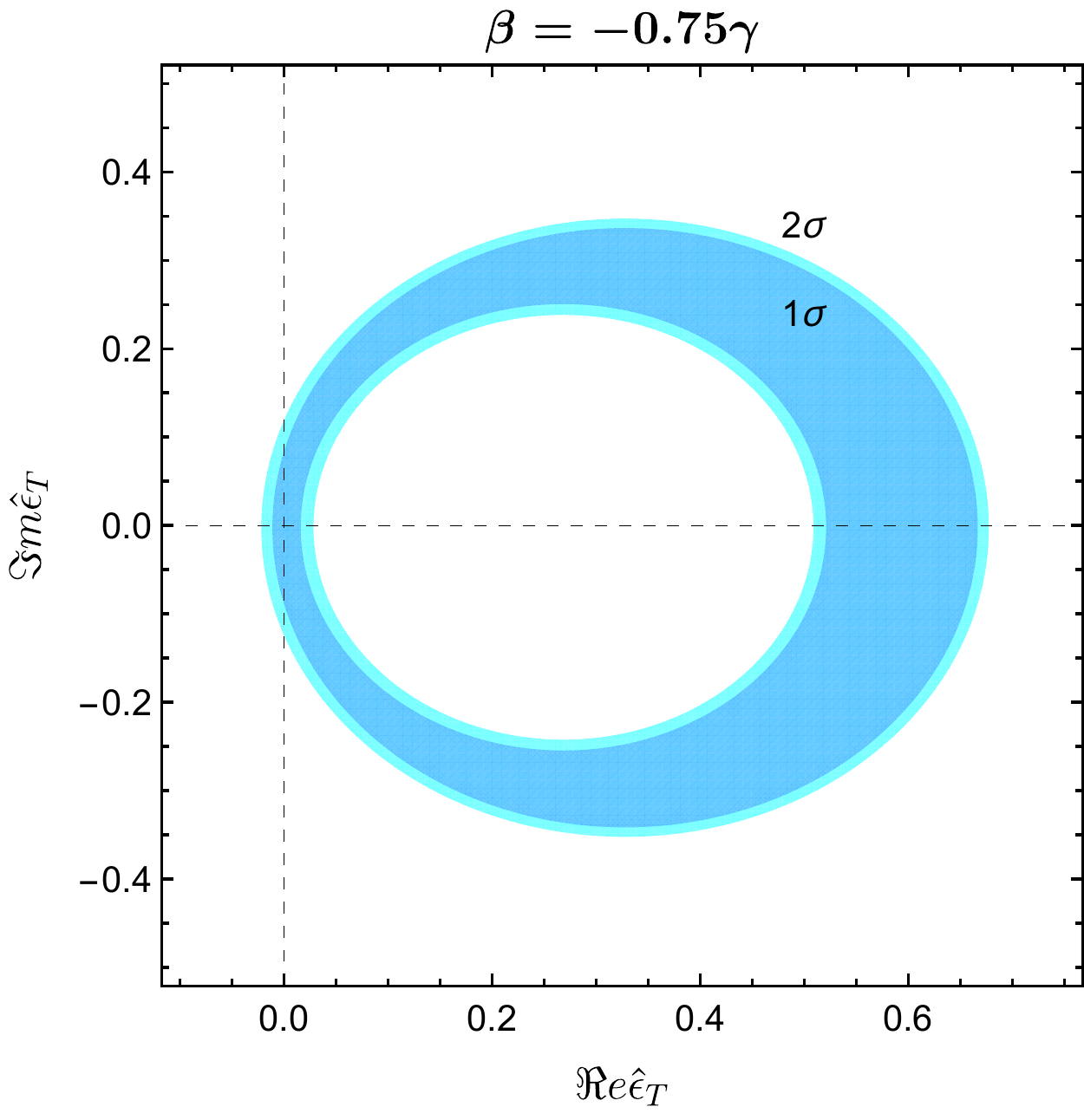}
  \caption{\label{fig:slimit} \small Constraints on $\hat{\epsilon}_S$ for $\hat{\epsilon}_T=0$ (left) as well as on $\hat{\epsilon}_T$ for $\hat{\epsilon}_S=0$ in both the $\beta=+0.75\gamma$ (middle) and $\beta=-0.75\gamma$ (right) cases, from the branching ratio $\mathcal B(\tau^-\to K_S\pi^-\nu_\tau)$ by varying it at both $1\sigma$ and $2\sigma$ intervals.}
\end{figure}

We now apply the observable $\Delta$ to put constrains on the non-standard scalar and tensor interactions. Since the effective couplings $\hat{\epsilon}_S$ and $\hat{\epsilon}_T$ are both considered to be complex, there are four degrees of freedom, $\Re e\hat{\epsilon}_S$, $\Im m\hat{\epsilon}_S$, $\Re e\hat{\epsilon}_T$ and $\Im m\hat{\epsilon}_T$, at our disposal. Combining our prediction for the branching ratio,
\begin{equation}
\mathcal{B}(\tau^-\to K_S\pi^-\nu_\tau)_{\text{SM}}=(0.421\pm0.022)\%\,,
\end{equation}
with the experimental result measured by the Belle collaboration~\cite{Ryu:2014vpc},
\begin{equation}
\mathcal{B}(\tau^-\to K_S\pi^-\nu_\tau)_{\text{Exp}}=(0.416\pm0.001(\text{stat})\pm0.008(\text{syst}))\%\,,
\end{equation}
we obtain the allowed regions, $\Delta\in[-0.07,0.05]$ and $\Delta\in[-0.12,0.10]$, by varying both the theoretical and experimental uncertainties at $1\sigma$ and $2\sigma$, respectively. To set bounds on one of the couplings $\hat{\epsilon}_S$ and $\hat{\epsilon}_T$, we shall assume the other to be zero, and our final results are shown in Fig.~\ref{fig:slimit} for both the $\beta=+0.75\gamma$ and $\beta=-0.75\gamma$ cases. It can be clearly seen that, under the constraint from the observable $\Delta$, the allowed region of $\hat{\epsilon}_S$ is larger than that of $\hat{\epsilon}_T$, which is consistent with our previous observation that a slightly larger sensitivity to the tensor than to the scalar contribution is preferred by the branching ratio. It is also observed that, while the imaginary parts of the allowed regions of both $\hat{\epsilon}_S$ and $\hat{\epsilon}_T$ are nearly symmetric about the axes, the real parts are not, but with the preference $\Re e\hat{\epsilon}_S<0$ and $\Re e\hat{\epsilon}_T>0$.

\begin{figure}[ht]
	\centering
	\includegraphics[width=3.15in]{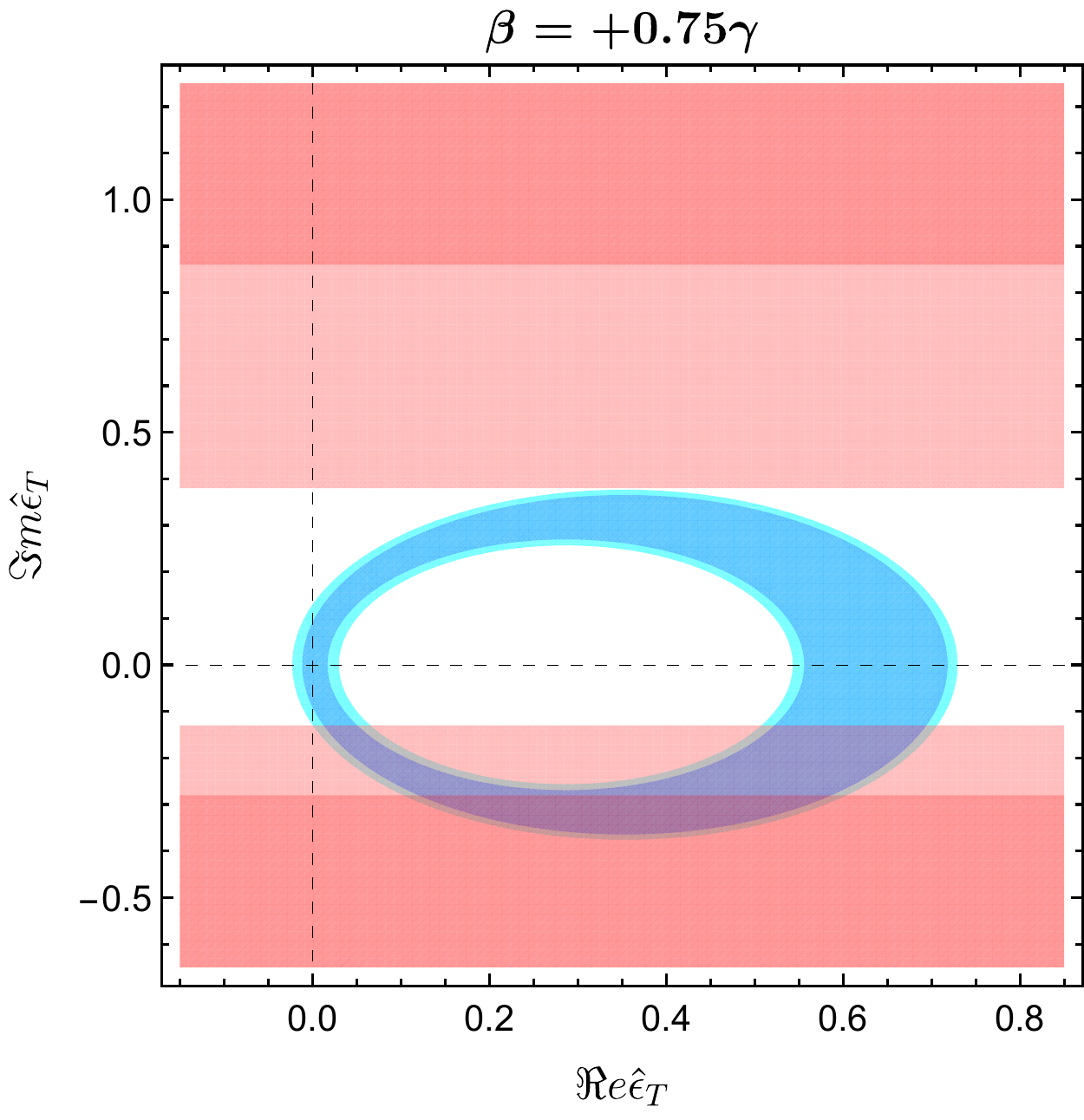}
	\hspace{0.15in}
	\includegraphics[width=3.15in]{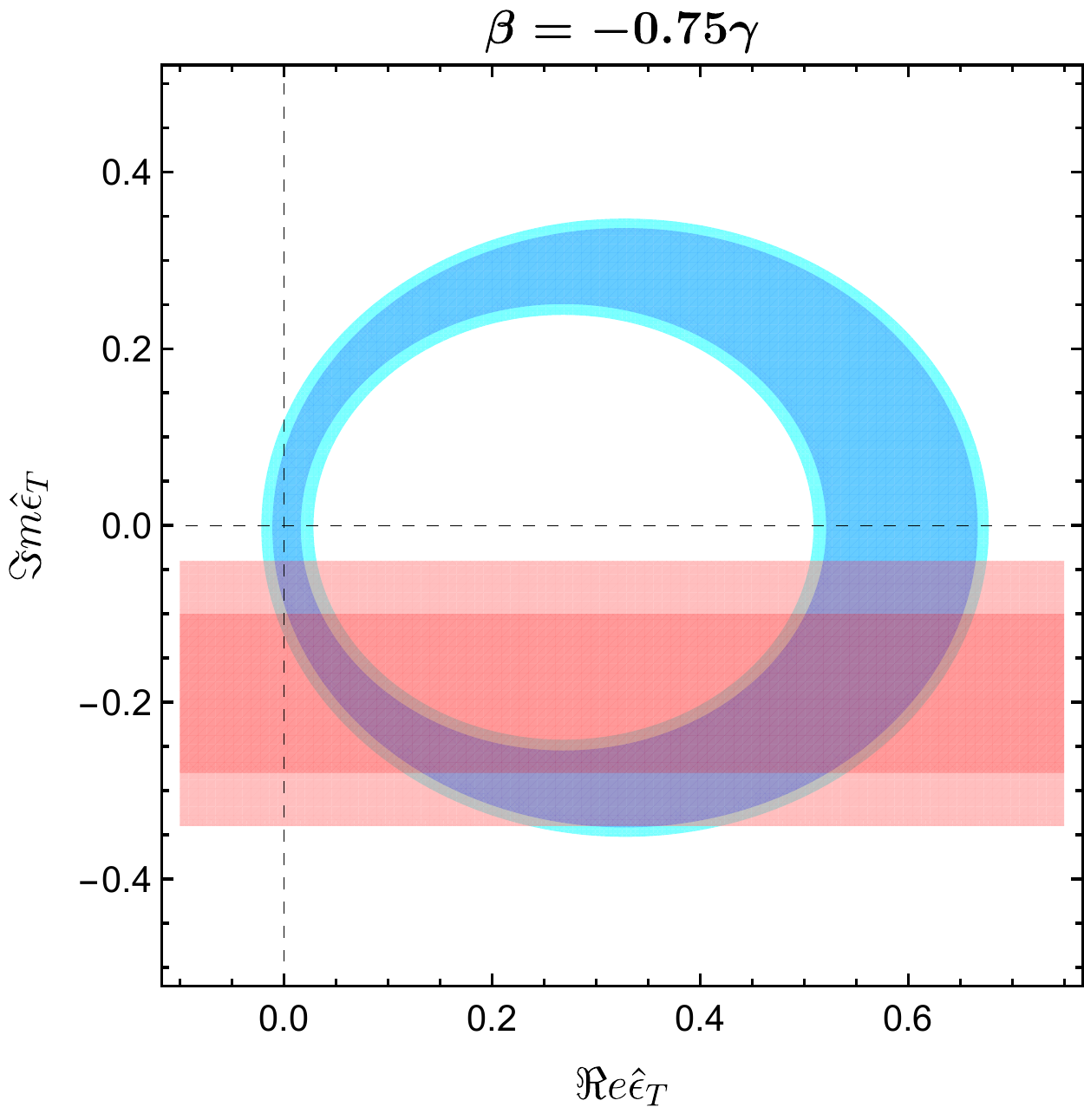}
	\caption{\label{fig:climit} \small Constraints on $\hat{\epsilon}_T$ from the branching ratio $\mathcal B(\tau^-\to K_S\pi^-\nu_\tau)$ (blue and cyan regions obtained at $1\sigma$ and $2\sigma$, respectively) as well as the decay-rate asymmetry $\mathcal A$ (red and pink regions at $1\sigma$ and $2\sigma$, respectively), in both the $\beta=+0.75\gamma$ (left) and $\beta=-0.75\gamma$ (right) cases.}
\end{figure}

As only the interference between vector and tensor operators can provide a potential NP explanation of the CP anomaly observed in $\tau\to K_S\pi\nu_\tau$ decays~\cite{Devi:2013gya,Cirigliano:2017tqn}, we now focus on the tensor coupling $\hat{\epsilon}_T$. To check if the region of $\hat{\epsilon}_T$ allowed by the branching ratio is compatible with that required by the CP asymmetry, we now add the constraint from the measured decay-rate asymmetry $\mathcal A$ by the BaBar collaboration~\cite{BABAR:2011aa}, and our final results are shown in Fig.~\ref{fig:climit} for both the $\beta=+0.75\gamma$ and $\beta=-0.75\gamma$ cases. One can see that, in both of these two cases, there exist common regions of the tensor coupling $\hat{\epsilon}_T$ that can accommodate both the branching ratio $\mathcal B(\tau^-\to K_S\pi^-\nu_\tau)$ and the CP asymmetry $A_{CP}(\tau\to K_S\pi\nu_\tau)$ simultaneously, even at the $1\sigma$ level. It is also observed that the $\beta=-0.75\gamma$ case is even preferred, in which a larger allowed region of $\hat\epsilon_T$ is obtained due to the slightly larger phase difference between the vector and tensor form factors, as mentioned already  the last section.

\subsection{Results in the scalar LQ scenario}

In the scalar LQ scenario, due to the specific relation $C_{S}(\mu_\phi)=-4\,C_{T}(\mu_\phi)$ at the matching scale $\mu_\phi=M_\phi$, we are actually left with only one effective coupling $\hat C_T$, and more severe constraint on it is therefore expected than in the model-independent case.

\begin{figure}[ht]
	\centering
	\includegraphics[width=3.15in]{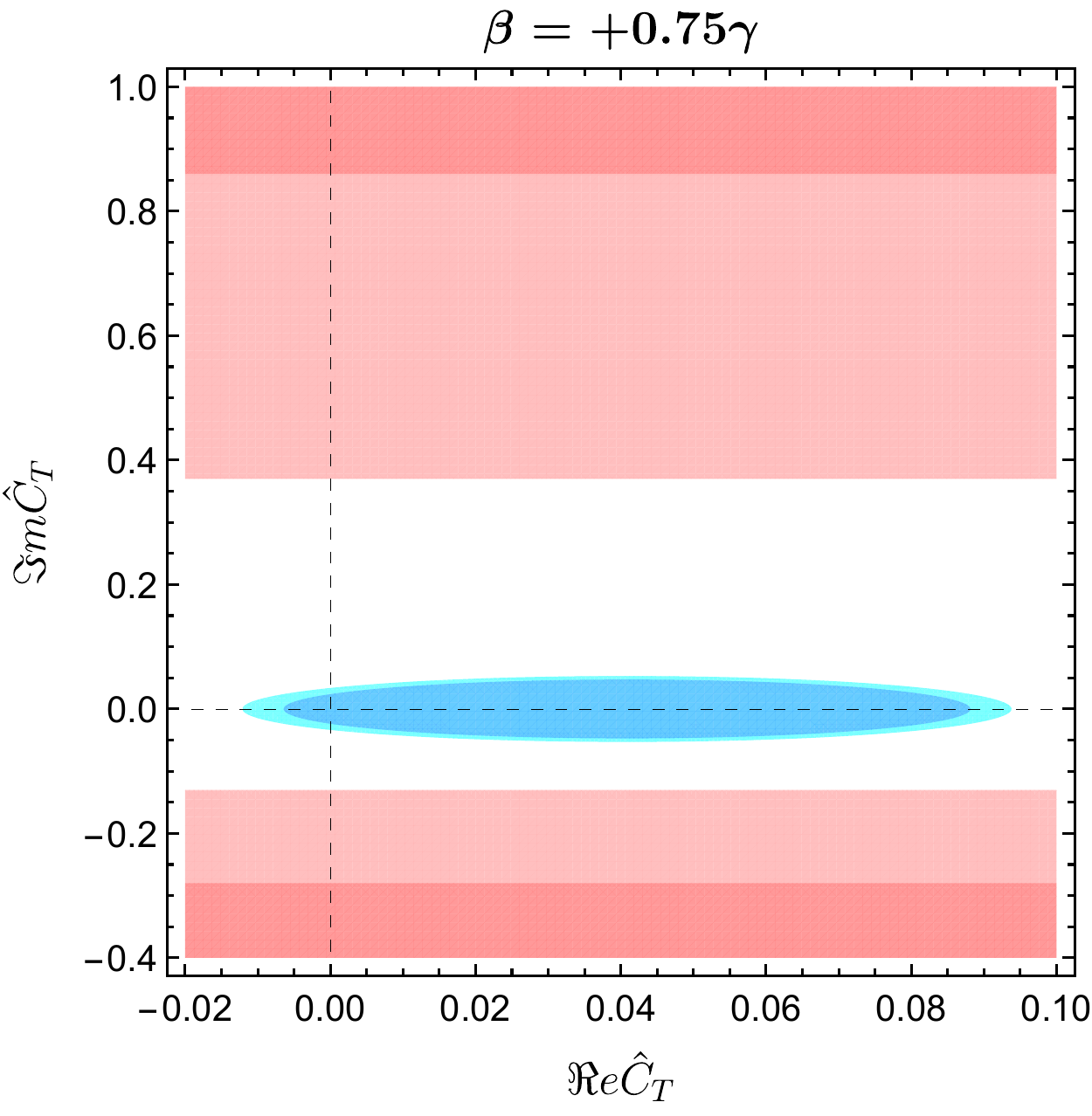}
	\hspace{0.15in}
	\includegraphics[width=3.15in]{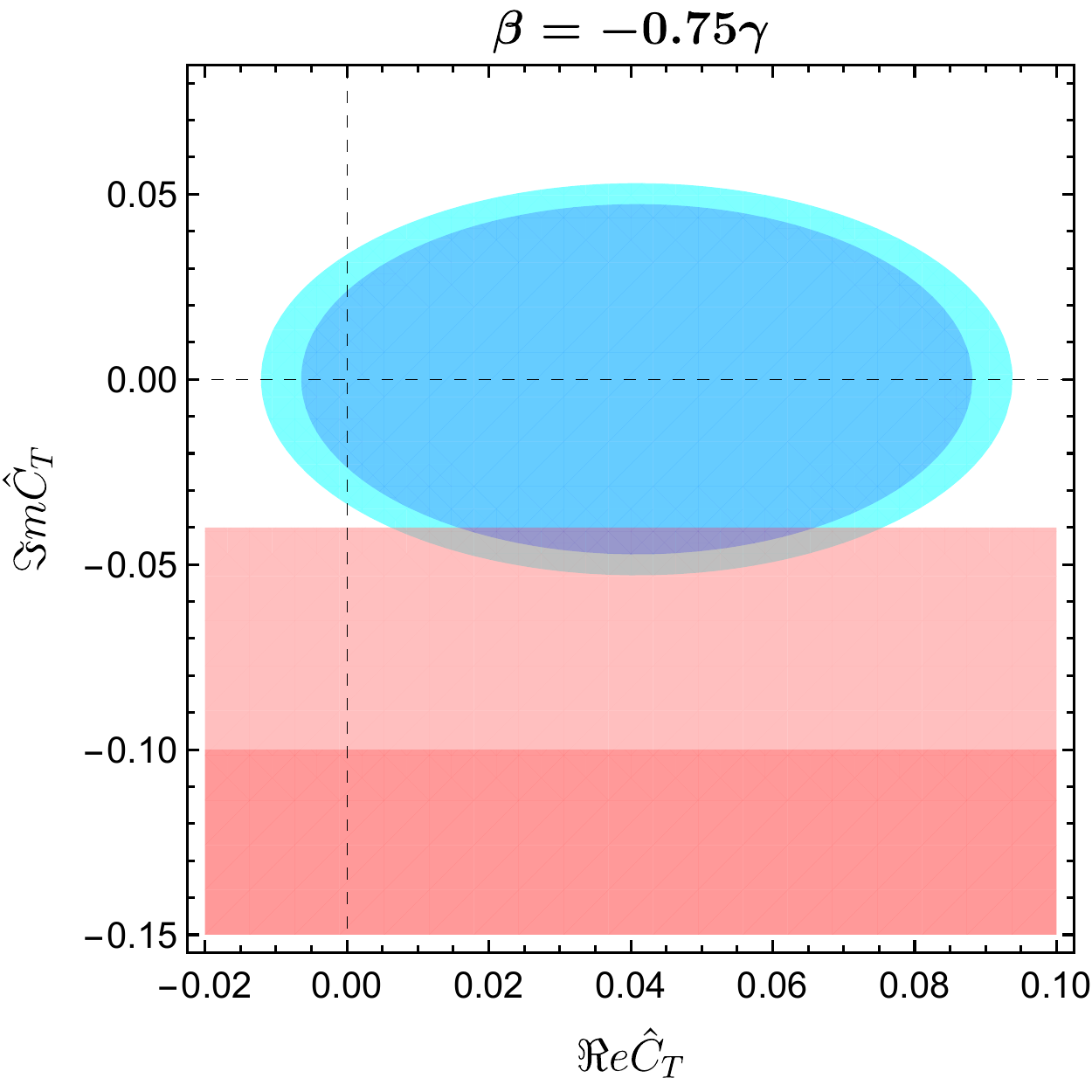}
	\caption{\label{fig:LQ} \small Constraints on $\hat{C}_T$ in the scalar LQ scenario. The other captions are the same as in Fig.~\ref{fig:climit}. }
\end{figure}

Referring to Eq.~\eqref{eq:WCLQ_low} and by fixing $M_\phi=1\,\text{TeV}$ and $\mu_\tau=m_\tau$, one obtains $\hat{C}_S\simeq-9.84\,\hat{C}_T$ at the $\mu_{\tau}$ scale. This implies that the scalar contribution is enhanced relative to that from the tensor operator in such a scenario. Under the constraints from the branching ratio $\mathcal B(\tau^-\to K_S\pi^-\nu_\tau)$ and the CP asymmetry $A_{CP}(\tau\to K_S\pi\nu_\tau)$, our final allowed regions of $\hat C_T$ are shown in Fig.~\ref{fig:LQ}. One can see that there is no common region allowed simultaneously by these two observables at the $1\sigma$ level, and only a small region is allowed in the $\beta=-0.75\gamma$ case at the $2\sigma$ level. This implies that the scalar LQ scenario can hardly account for the observed CP anomaly under the constraint from the measured branching ratio, except for the marginal region obtained in the $\beta=-0.75\gamma$ case at the $2\sigma$ level.

\subsection{Constraints from other observables and processes}

It should be noted that in both of the two scenarios discussed above, we have assumed that the non-standard scalar and tensor operators contribute only to the $\tau\to K_S\pi\nu_\tau$ decays, and only the branching ratio $\mathcal B(\tau^-\to K_S\pi^-\nu_\tau)$ and the CP asymmetry $A_{CP}(\tau\to K_S\pi\nu_\tau)$ have been considered to constrain the corresponding effective couplings. We now discuss the constraints on these non-standard interactions from other observables and processes.

As pointed out already in Refs.~\cite{Rendon:2019awg,Roig:2019rwf}, the $\tau^-\to K_S\pi^-\nu_\tau$ decay spectrum measured by the Belle collaboration~\cite{Epifanov:2007rf} can also provide very complementary constraints on these non-standard interactions. Under the combined constraints from the branching ratio and the decay spectrum of this decay, the best fit values $\hat{\epsilon}_{S}=(1.3\pm0.9)\times10^{-2}$ and $\hat{\epsilon}_{T}=(0.7\pm1.0)\times10^{-2}$ have been obtained in Ref.~\cite{Rendon:2019awg}. Assuming that $\Re e\hat{\epsilon}_{S(T)}\sim\Im m\hat{\epsilon}_{S(T)}\sim\hat{\epsilon}_{S(T)}$\footnote{The couplings $\hat{\epsilon}_{S}$ and $\hat{\epsilon}_{T}$ are assumed to be real in Ref.~\cite{Rendon:2019awg}. As the modulus of the tensor form factor is almost unaffected by the inelastic effect and is quite consistent with that obtained with the Omn\`es factor $\Omega(s)$ (see Figs.~\ref{fig:positive} and \ref{fig:negative}), similar numerical results as that obtained in Ref.~\cite{Rendon:2019awg} are expected, even though a different phase, $\delta_T(s)=\delta_+(s)$ in Ref.~\cite{Rendon:2019awg} versus Eq.~\eqref{eq:nT} in this work, has been adopted for the tensor form factor.}, one can see that these values are smaller by at least one order of magnitude than our results obtained under the constraint from only the branching ratio (see Fig.~\ref{fig:slimit}). This implies that, once the combined constraints from the branching ratio and the decay spectrum are taken into account, the allowed value of the tensor coupling $\hat{\epsilon}_{T}$ will be insufficient to explain the CP anomaly, which demands that $\Im m\hat{\epsilon}_{T}$ should be of the order $\mathcal{O}(10^{-1})$ at least (see Figs.~\ref{fig:climit} and \ref{fig:LQ})~\cite{Cirigliano:2017tqn}.

If the lepton-flavour university (LFU) is further assumed\footnote{Although the LFU is hinted to be violated by the current data on $B$-meson decays (see Refs.~\cite{HFLAV:2019,Bifani:2018zmi,Ciezarek:2017yzh,Li:2018lxi} and references therein for recent reviews), there exists up to now no compelling evidence for its violation in the strangeness sector~\cite{Tanabashi:2018oca}. Actually, the charged-current anomalies observed in semi-leptonic $B$ decays become already the least compelling hints for the LFU violation by the latest Belle data~\cite{Belle:2019rba}.}, the effective operators given by Eq.~\eqref{eq:Efective_Lagrangian} but with the $\tau$ lepton replaced by the electron and muon flavours would also contribute to other strangeness-changing processes. In this case, our bounds on the non-standard scalar and tensor couplings would be not competitive with that obtained from the (semi-)leptonic kaon~\cite{Gonzalez-Alonso:2016etj} and hyperon~\cite{Chang:2014iba} decays, due to the larger systematic theory uncertainty inherent to the current framework for hadronic $\tau$ decays, especially in the inelastic region. For example, the global fit results,  $\hat{\epsilon}_{S}=(-3.9\pm4.9)\times10^{-4}$ and $\hat{\epsilon}_{T}=(0.5\pm5.2)\times10^{-3}$, from the (semi-)leptonic kaon decays~\cite{Gonzalez-Alonso:2016etj} are already much stronger than our bounds shown in Figs.~\ref{fig:climit} and \ref{fig:LQ}. 

It is also noted that, unless some NP between the electroweak and the low-energy scale is assumed, the effective Lagrangian specified by Eq.~\eqref{eq:Efective_Lagrangian} comes generally from an $SU(2)$-invariant form~\cite{Buchmuller:1985jz,Grzadkowski:2010es,Alonso:2014csa}. Thus, the demand of $SU(2)$ invariance of the weak interactions naturally relates the tensor operator relevant for $\tau\to K_S\pi\nu_\tau$ to the neutral-current tensor operator relevant for the neutron EDM and the $D-\bar{D}$ mixing, as pointed out already in Ref.~\cite{Cirigliano:2017tqn}. This also brings the tensor coupling required by the CP asymmetry $A_{CP}(\tau\to K_S\pi\nu_\tau)$ to be already in conflict with the bounds from these two observables, leading to the claim that it is extremely difficult to explain the CP anomaly in terms of ultra-violet complete NP scenarios~\cite{Cirigliano:2017tqn}. 

Based on the above observations, we conclude therefore that it is quite difficult to explain the CP anomaly within the two frameworks considered here, as claimed already in Refs.~\cite{Cirigliano:2017tqn,Rendon:2019awg}.

\section{Conclusion}
\label{sec:conclusion}

In this paper, motivated by the $2.8\sigma$ discrepancy observed between the BaBar measurement and the SM prediction of the CP asymmetry in $\tau\to K_S\pi\nu_\tau$ decays, as well as the prospects of future measurements at Belle II, we have studied this observable within the model-independent low-energy effective theory framework and in the scalar LQ scenario, both of which contain a non-standard tensor operator that is necessary to produce a non-vanishing direct CP asymmetry in the decays considered. Our main conclusions are summarized as follows:
\begin{itemize}
\item By employing the reciprocal basis, which is found to be most convenient when a $K_{S}$ or $K_{L}$ is involved in the final state, we have reproduced the known CP asymmetry due to $K^0 -\bar{K}^0$ mixing, as predicted firstly by Bigi and Sanda~\cite{Bigi:2005ts} but with a sign mistake and then corrected by Grossman and Nir~\cite{Grossman:2011zk}.

\item As the $K\pi$ tensor form factor plays a crucial role in generating a non-zero direct CP asymmetry that can arise only from the interference of vector and tensor operators, we have presented a new calculation of this form factor in the context of $\chi$PT with tensor sources and R$\chi$T with both $K^\ast(892)$ and $K^\ast(1410)$ included. For these spin-1 vector resonances, we have used the more conventional vector representation instead of the description based on anti-symmetric tensor fields. A once-subtracted dispersive representation of this form factor has also been presented, which naturally fulfills the requirements of unitarity and analyticity. Furthermore, our estimate of the relation between the two weight parameters, $\beta\simeq -0.75\gamma$, gives a support for the assumption made in Ref.~\cite{Cirigliano:2017tqn} that the inelastic contributions to the phases of vector and tensor form factors are of similar size but potentially opposite in sign. 

\item Adopting the three-times subtracted (for the vector form factor) and the coupled-channel (for the scalar form factor) dispersive representations, together with our result of the tensor form factor, we have performed a detailed analysis of the $\tau\to K_S\pi\nu_\tau$ decays within the two scenarios mentioned above. It is observed that the CP anomaly can be accommodated in the model-independent framework, even at the $1\sigma$ level, together with the constraint from the branching ratio of $\tau^-\to K_S\pi^-\nu_\tau$ decay. In the LQ scenario, however, this anomaly can be marginally reconciled only at the $2\sigma$ level, due to the specific relation between the scalar and tensor operators. Once the combined constraints from the branching ratio and the decay spectrum of this decay are taken into account, these two possibilities are however both excluded, even without exploiting further the stronger bounds from the (semi-)leptonic kaon decays~\cite{Gonzalez-Alonso:2016etj} under the assumption of lepton-flavour universality, as well as from the neutron EDM and $D-\bar{D}$ mixing under the assumption of $SU(2)$ invariance of the weak interactions~\cite{Cirigliano:2017tqn}. It is therefore difficult to explain such a CP anomaly within the frameworks considered here. 
\end{itemize}

As both the theoretical predictions and the experimental measurements are still plagued by large uncertainties, more refined studies, especially the information on the $K\pi$ tensor form factor in the inelastic region as well as the dedicated measurements of $\tau\to K_S\pi\nu_\tau$ decays from the Belle II collaboration~\cite{Kou:2018nap}, are expected.

\section*{Acknowledgements}

We are grateful to Toni Pich, Jorge Portoles and Pablo Roig for valuable information about the $K\pi$ form factors. This work is supported by the National Natural Science Foundation of China  under Grant Nos.~11675061, 11775092 and 11435003. X.L. is also supported in part by the Fundamental Research Funds for the Central Universities under Grant No.~CCNU18TS029.

\appendix
\renewcommand{\theequation}{A.\arabic{equation}}
\section*{Appendix: Input parameters}
\label{app:block}

In this appendix, for convenience, we collect in Table~\ref{tab:input} all the input parameters used throughout this paper. For further details, the readers are referred to the original references.

\begin{table}[ht]
\tabcolsep 0.045in
\renewcommand\arraystretch{1.7}
\begin{center}
	\caption{\label{tab:input} \small Summary of the input parameters used throughout this paper.}
	\vspace{0.18cm}
		\begin{tabular}{|c|c|c|c|c|c|}
			\hline\hline
			\multicolumn{6}{|l|}{QCD and electroweak parameters~\cite{Tanabashi:2018oca}}\\
			\hline
			$G_F[10^{-5}~\text{GeV}^{-2}]$ & $\alpha_s(M_Z)$ & $m_t~[\text{GeV}]$ & $m_b~[\text{GeV}]$ & $F_{\pi}~[\text{MeV}]$  & $F_K~[\text{MeV}]$\\
			\hline
			$1.1663787(6)$ & $0.1181(11)$ & $173.1$ & $4.18$ & $92.3(1)$  & $1.198F_\pi$\\
			\hline
			\multicolumn{6}{|l|}{Particle masses and $\tau$ lifetime~\cite{Tanabashi:2018oca}}\\
			\hline
			$m_{\tau}~[\text{MeV}]$ & $M_{K^0}~[\text{MeV}]$ & $M_{\pi^-}~[\text{MeV}]$ & \multicolumn{3}{c|}{$\tau_\tau~[10^{-15}~\text{s}]$}\\
			\hline
			$1776.86$ & $497.61$ & $139.57$ & \multicolumn{3}{c|}{$290.3$}\\
			\hline
			\multicolumn{6}{|l|}{Parameters in the $K\pi$ vector form factor with $s_{cut}=4~\text{GeV}^2$~\cite{Boito:2008fq}}\\
			\hline
			$m_{K^*}~[\text{MeV}]$ & $\gamma_{K^*}~[\text{MeV}]$ & $m_{K^{*\prime}}~[\text{MeV}]$ & $\gamma_{K^{*\prime}}~[\text{MeV}]$& \multicolumn{2}{c|}{$\gamma$}\\
			\hline
			$943.41\pm0.59$& $66.72\pm0.87$ & $1374\pm45$ & $240\pm131$ & \multicolumn{2}{c|}{$-0.039\pm0.020$}\\
			\hline
			\multicolumn{2}{|c|}{$M_{K^*}~[\text{MeV}]$}&\multicolumn{2}{|c|}{$\lambda_+^{\prime}$} & \multicolumn{2}{c|}{$\lambda_+^{\prime\prime}$ }\\
			\hline
			\multicolumn{2}{|c|}{$892.01\pm0.92$}&\multicolumn{2}{|c|}{$(24.66\pm0.77)\times10^{-3}$} & \multicolumn{2}{c|}{$(11.99\pm0.20)\times10^{-4}$}\\
			\hline
			\multicolumn{6}{|l|}{CP-violating parameters as well as the measured decay-rate asymmetry}\\
			\hline
			$\left|\eta_{+-}\right|\times10^{3}$~\cite{Tanabashi:2018oca}& $\phi_{+-}$~\cite{Tanabashi:2018oca}&$\Re e(\epsilon)\times10^{3}$~\cite{Tanabashi:2018oca}&$A^{\rm SM}_{CP}$~\cite{BABAR:2011aa}&\multicolumn{2}{|c|}{$\mathcal{A}$~\cite{BABAR:2011aa}}\\
			\hline
			$2.232\pm0.011$&$(43.51\pm0.05)^{\circ}$& $1.66\pm0.02$& $(0.36\pm0.01)\%$ &\multicolumn{2}{|c|}{$(-0.27\pm0.18\pm0.08)\%$}\\
			\hline
			\multicolumn{6}{|l|}{Other input parameters}\\
			\hline
			$M_{V^\prime}~[\text{MeV}]$~\cite{Jamin:2000wn} & $\Lambda_2~[\text{MeV}]$~\cite{Baum:2011rm} & $S_{\rm EW}$~\cite{Erler:2002mv} & $\left|V_{us}F_+(0)\right|$~\cite{Moulson:2017ive}&\multicolumn{2}{|c|}{$\mathcal{B}(\tau^-\to K_S\pi^-\nu_\tau)$~\cite{Ryu:2014vpc}}\\
			\hline
			$1440$ & $11.1(4)$ & $1.0201(3)$ & $0.21654(41)$ &\multicolumn{2}{|c|}{$(0.416\pm0.001\pm0.008)\%$}\\
			\hline \hline
		\end{tabular}
\end{center}
\end{table}

\bibliographystyle{JHEP}
\bibliography{reference}

\providecommand{\href}[2]{#2}\begingroup\raggedright\begin{thebibliography}{100}

\bibitem{Kobayashi:1973fv}
M.~Kobayashi and T.~Maskawa, {\it {CP Violation in the Renormalizable Theory of
  Weak Interaction}},  {\it Prog. Theor. Phys.} {\bf 49} (1973) 652--657.

\bibitem{Sakharov:1967dj}
A.~D. Sakharov, {\it {Violation of CP Invariance, C asymmetry, and baryon
  asymmetry of the universe}},  {\it Pisma Zh. Eksp. Teor. Fiz.} {\bf 5} (1967)
  32--35. [Usp. Fiz. Nauk161,no.5,61(1991)].

\bibitem{Huet:1994jb}
P.~Huet and E.~Sather, {\it {Electroweak baryogenesis and standard model CP
  violation}},  {\it Phys. Rev.} {\bf D51} (1995) 379--394,
  [\href{http://arxiv.org/abs/hep-ph/9404302}{{\tt hep-ph/9404302}}].

\bibitem{Cohen:1993nk}
A.~G. Cohen, D.~B. Kaplan, and A.~E. Nelson, {\it {Progress in electroweak
  baryogenesis}},  {\it Ann. Rev. Nucl. Part. Sci.} {\bf 43} (1993) 27--70,
  [\href{http://arxiv.org/abs/hep-ph/9302210}{{\tt hep-ph/9302210}}].

\bibitem{Riotto:1999yt}
A.~Riotto and M.~Trodden, {\it {Recent progress in baryogenesis}},  {\it Ann.
  Rev. Nucl. Part. Sci.} {\bf 49} (1999) 35--75,
  [\href{http://arxiv.org/abs/hep-ph/9901362}{{\tt hep-ph/9901362}}].

\bibitem{Pich:2013lsa}
A.~Pich, {\it {Precision Tau Physics}},  {\it Prog. Part. Nucl. Phys.} {\bf 75}
  (2014) 41--85, [\href{http://arxiv.org/abs/1310.7922}{{\tt
  arXiv:1310.7922}}].

\bibitem{Davier:2005xq}
M.~Davier, A.~Hocker, and Z.~Zhang, {\it {The Physics of hadronic tau decays}},
   {\it Rev. Mod. Phys.} {\bf 78} (2006) 1043--1109,
  [\href{http://arxiv.org/abs/hep-ph/0507078}{{\tt hep-ph/0507078}}].

\bibitem{Bigi:2012km}
I.~I. Bigi, {\it {Probing CP Violation in $\tau ^- \to \nu (K\pi/K2\pi / 3K/
  K3\pi)^- $ Decays}},  \href{http://arxiv.org/abs/1204.5817}{{\tt
  arXiv:1204.5817}}.

\bibitem{Bigi:2012kz}
I.~I. Bigi, {\it {CP Violation in $\tau$ Decays at SuperB $\&$ Super-Belle II
  Experiments - like Finding Signs of Dark Matter}},  {\it Nucl. Phys. Proc.
  Suppl.} {\bf 253-255} (2014) 91--94,
  [\href{http://arxiv.org/abs/1210.2968}{{\tt arXiv:1210.2968}}].

\bibitem{Kiers:2012fy}
K.~Kiers, {\it {CP violation in hadronic $\tau$ decays}},  {\it Nucl. Phys.
  Proc. Suppl.} {\bf 253-255} (2014) 95--98,
  [\href{http://arxiv.org/abs/1212.6921}{{\tt arXiv:1212.6921}}].

\bibitem{Bonvicini:2001xz}
{\bf CLEO} Collaboration, G.~Bonvicini et~al., {\it {Search for CP violation in
  tau $\to$ K pi tau-neutrino decays}},  {\it Phys. Rev. Lett.} {\bf 88} (2002)
  111803, [\href{http://arxiv.org/abs/hep-ex/0111095}{{\tt hep-ex/0111095}}].

\bibitem{Bischofberger:2011pw}
{\bf Belle} Collaboration, M.~Bischofberger et~al., {\it {Search for CP
  violation in $\tau \to K^0_S \pi \nu_\tau$ decays at Belle}},  {\it Phys.
  Rev. Lett.} {\bf 107} (2011) 131801,
  [\href{http://arxiv.org/abs/1101.0349}{{\tt arXiv:1101.0349}}].

\bibitem{BABAR:2011aa}
{\bf BaBar} Collaboration, J.~P. Lees et~al., {\it {Search for CP Violation in
  the Decay $\tau^- \to \pi^- K^0_S (>= 0 \pi^0) \nu_\tau$}},  {\it Phys. Rev.}
  {\bf D85} (2012) 031102, [\href{http://arxiv.org/abs/1109.1527}{{\tt
  arXiv:1109.1527}}]. [Erratum: Phys. Rev.D85,099904(2012)].

\bibitem{Delepine:2005tw}
D.~Delepine, G.~Lopez~Castro, and L.~T. Lopez~Lozano, {\it {CP violation in
  semileptonic tau lepton decays}},  {\it Phys. Rev.} {\bf D72} (2005) 033009,
  [\href{http://arxiv.org/abs/hep-ph/0503090}{{\tt hep-ph/0503090}}].

\bibitem{Christenson:1964fg}
J.~H. Christenson, J.~W. Cronin, V.~L. Fitch, and R.~Turlay, {\it {Evidence for
  the $2\pi$ Decay of the $K_2^0$ Meson}},  {\it Phys. Rev. Lett.} {\bf 13}
  (1964) 138--140.

\bibitem{Tanabashi:2018oca}
{\bf Particle Data Group} Collaboration, M.~Tanabashi et~al., {\it {Review of
  Particle Physics}},  {\it Phys. Rev.} {\bf D98} (2018), no.~3 030001.

\bibitem{Bigi:2005ts}
I.~I. Bigi and A.~I. Sanda, {\it {A `Known' CP asymmetry in tau decays}},  {\it
  Phys. Lett.} {\bf B625} (2005) 47--52,
  [\href{http://arxiv.org/abs/hep-ph/0506037}{{\tt hep-ph/0506037}}].

\bibitem{Grossman:2011zk}
Y.~Grossman and Y.~Nir, {\it {CP Violation in $\tau\to\nu\pi K_S$ and $D\to\pi
  K_S$: The Importance of $K_S-K_L$ Interference}},  {\it JHEP} {\bf 04} (2012)
  002, [\href{http://arxiv.org/abs/1110.3790}{{\tt arXiv:1110.3790}}].

\bibitem{Calderon:2007rg}
G.~Calderon, D.~Delepine, and G.~L. Castro, {\it {Is there a paradox in CP
  asymmetries of tau+- $\to$ K(L,S)pi+- nu decays?}},  {\it Phys. Rev.} {\bf
  D75} (2007) 076001, [\href{http://arxiv.org/abs/hep-ph/0702282}{{\tt
  hep-ph/0702282}}].

\bibitem{Kou:2018nap}
{\bf Belle-II} Collaboration, W.~Altmannshofer et~al., {\it {The Belle II
  Physics Book}},  \href{http://arxiv.org/abs/1808.10567}{{\tt
  arXiv:1808.10567}}.

\bibitem{Devi:2013gya}
H.~Z. Devi, L.~Dhargyal, and N.~Sinha, {\it {Can the observed CP asymmetry in
  $\tau \to K\pi\nu_\tau$ be due to nonstandard tensor interactions?}},  {\it
  Phys. Rev.} {\bf D90} (2014), no.~1 013016,
  [\href{http://arxiv.org/abs/1308.4383}{{\tt arXiv:1308.4383}}].

\bibitem{Dhargyal:2016kwp}
L.~Dhargyal, {\it {Full angular spectrum analysis of tensor current
  contribution to $A_{cp}(\tau \rightarrow K_{s} \pi \nu_{\tau})$}},  {\it
  LHEP} {\bf 1} (2018), no.~3 9--14,
  [\href{http://arxiv.org/abs/1605.00629}{{\tt arXiv:1605.00629}}].

\bibitem{Dhargyal:2016jgo}
L.~Dhargyal, {\it {New tensor interaction as the source of the observed CP
  asymmetry in $\tau \to K_{S}\pi\nu_{\tau}$}},  {\it Springer Proc. Phys.}
  {\bf 203} (2018) 329--331, [\href{http://arxiv.org/abs/1610.06293}{{\tt
  arXiv:1610.06293}}].

\bibitem{Cirigliano:2017tqn}
V.~Cirigliano, A.~Crivellin, and M.~Hoferichter, {\it {A no-go theorem for
  non-standard explanations of the $\tau\to K_S\pi\nu_\tau$ CP asymmetry}},
  {\it Phys. Rev. Lett.} {\bf 120} (2018), no.~14 141803,
  [\href{http://arxiv.org/abs/1712.06595}{{\tt arXiv:1712.06595}}].

\bibitem{Rendon:2019awg}
J.~Rendón, P.~Roig, and G.~Toledo~Sánchez, {\it {Effective-field theory
  analysis of the $\tau^{-}\rightarrow (K \pi)^{-}\nu_{\tau}$ decays}},  {\it
  Phys. Rev.} {\bf D99} (2019), no.~9 093005,
  [\href{http://arxiv.org/abs/1902.08143}{{\tt arXiv:1902.08143}}].

\bibitem{Dighe:2019odu}
A.~Dighe, S.~Ghosh, G.~Kumar, and T.~S. Roy, {\it {Tensors for tending to
  tensions in $ \tau $ decays}},  \href{http://arxiv.org/abs/1902.09561}{{\tt
  arXiv:1902.09561}}.

\bibitem{Antonelli:2013usa}
M.~Antonelli, V.~Cirigliano, A.~Lusiani, and E.~Passemar, {\it {Predicting the
  $\tau$ strange branching ratios and implications for $V_{us}$}},  {\it JHEP}
  {\bf 10} (2013) 070, [\href{http://arxiv.org/abs/1304.8134}{{\tt
  arXiv:1304.8134}}].

\bibitem{GodinaNava:1995jb}
J.~J. Godina~Nava and G.~Lopez~Castro, {\it {Tensor interactions and tau
  decays}},  {\it Phys. Rev.} {\bf D52} (1995) 2850--2854,
  [\href{http://arxiv.org/abs/hep-ph/9506330}{{\tt hep-ph/9506330}}].

\bibitem{Delepine:2006fv}
D.~Delepine, G.~Faisl, S.~Khalil, and G.~L. Castro, {\it {Supersymmetry and CP
  violation in $|\Delta S| = 1$ tau-decays}},  {\it Phys. Rev.} {\bf D74}
  (2006) 056004, [\href{http://arxiv.org/abs/hep-ph/0608008}{{\tt
  hep-ph/0608008}}].

\bibitem{Watson:1954uc}
K.~M. Watson, {\it {Some general relations between the photoproduction and
  scattering of pi mesons}},  {\it Phys. Rev.} {\bf 95} (1954) 228--236.

\bibitem{Epifanov:2007rf}
{\bf Belle} Collaboration, D.~Epifanov et~al., {\it {Study of $\tau^-\to
  K_S\pi^-\nu_\tau$ decay at Belle}},  {\it Phys. Lett.} {\bf B654} (2007)
  65--73, [\href{http://arxiv.org/abs/0706.2231}{{\tt arXiv:0706.2231}}].

\bibitem{Paramesvaran:2009ec}
{\bf BaBar} Collaboration, S.~Paramesvaran, {\it {Selected topics in tau
  physics from BaBar}},  in {\it {Particles and fields. Proceedings, Meeting of
  the Division of the American Physical Society, DPF 2009, Detroit, USA, July
  26-31, 2009}}, 2009.
\newblock \href{http://arxiv.org/abs/0910.2884}{{\tt arXiv:0910.2884}}.

\bibitem{Finkemeier:1995sr}
M.~Finkemeier and E.~Mirkes, {\it {Tau decays into kaons}},  {\it Z. Phys.}
  {\bf C69} (1996) 243--252, [\href{http://arxiv.org/abs/hep-ph/9503474}{{\tt
  hep-ph/9503474}}].

\bibitem{Finkemeier:1996dh}
M.~Finkemeier and E.~Mirkes, {\it {The Scalar contribution to tau $\to$ k pi
  tau-neutrino}},  {\it Z. Phys.} {\bf C72} (1996) 619--626,
  [\href{http://arxiv.org/abs/hep-ph/9601275}{{\tt hep-ph/9601275}}].

\bibitem{Jamin:2006tk}
M.~Jamin, A.~Pich, and J.~Portoles, {\it {Spectral distribution for the decay
  tau $\to$ nu(tau) K pi}},  {\it Phys. Lett.} {\bf B640} (2006) 176--181,
  [\href{http://arxiv.org/abs/hep-ph/0605096}{{\tt hep-ph/0605096}}].

\bibitem{Jamin:2008qg}
M.~Jamin, A.~Pich, and J.~Portoles, {\it {What can be learned from the Belle
  spectrum for the decay $\tau^- \to \nu_{\tau} K_S \pi^-$}},  {\it Phys.
  Lett.} {\bf B664} (2008) 78--83, [\href{http://arxiv.org/abs/0803.1786}{{\tt
  arXiv:0803.1786}}].

\bibitem{Boito:2008fq}
D.~R. Boito, R.~Escribano, and M.~Jamin, {\it { $K\pi$ vector form-factor,
  dispersive constraints and $\tau \to\nu_\tau K \pi$ decays}},  {\it Eur.
  Phys. J.} {\bf C59} (2009) 821--829,
  [\href{http://arxiv.org/abs/0807.4883}{{\tt arXiv:0807.4883}}].

\bibitem{Boito:2010me}
D.~R. Boito, R.~Escribano, and M.~Jamin, {\it { $K\pi$ vector form factor
  constrained by $\tau \to K\pi \nu_\tau$ and $K_{\ell3}$ decays}},  {\it JHEP}
  {\bf 09} (2010) 031, [\href{http://arxiv.org/abs/1007.1858}{{\tt
  arXiv:1007.1858}}].

\bibitem{Jamin:2000wn}
M.~Jamin, J.~A. Oller, and A.~Pich, {\it {S wave K pi scattering in chiral
  perturbation theory with resonances}},  {\it Nucl. Phys.} {\bf B587} (2000)
  331--362, [\href{http://arxiv.org/abs/hep-ph/0006045}{{\tt hep-ph/0006045}}].

\bibitem{Jamin:2001zq}
M.~Jamin, J.~A. Oller, and A.~Pich, {\it {Strangeness changing scalar
  form-factors}},  {\it Nucl. Phys.} {\bf B622} (2002) 279--308,
  [\href{http://arxiv.org/abs/hep-ph/0110193}{{\tt hep-ph/0110193}}].

\bibitem{Jamin:2006tj}
M.~Jamin, J.~A. Oller, and A.~Pich, {\it {Scalar K pi form factor and light
  quark masses}},  {\it Phys. Rev.} {\bf D74} (2006) 074009,
  [\href{http://arxiv.org/abs/hep-ph/0605095}{{\tt hep-ph/0605095}}].

\bibitem{Escribano:2014joa}
R.~Escribano, S.~González-Solís, M.~Jamin, and P.~Roig, {\it {Combined
  analysis of the decays $\tau^{-} \to K_{S}\pi^{-} \nu_{\tau}$ and $\tau^{-}
  \to K^{-}\eta\nu_{\tau}$}},  {\it JHEP} {\bf 09} (2014) 042,
  [\href{http://arxiv.org/abs/1407.6590}{{\tt arXiv:1407.6590}}].

\bibitem{Moussallam:2007qc}
B.~Moussallam, {\it {Analyticity constraints on the strangeness changing vector
  current and applications to tau $\to$ K pi nu(tau), tau $\to$ K pi pi
  nu(tau)}},  {\it Eur. Phys. J.} {\bf C53} (2008) 401--412,
  [\href{http://arxiv.org/abs/0710.0548}{{\tt arXiv:0710.0548}}].

\bibitem{Bernard:2013jxa}
V.~Bernard, {\it {First determination of $f_+(0) |V_{us}|$ from a combined
  analysis of $\tau\to K\pi \nu_\tau$ decay and $\pi K$ scattering with
  constraints from $K_{\ell3}$ decays}},  {\it JHEP} {\bf 06} (2014) 082,
  [\href{http://arxiv.org/abs/1311.2569}{{\tt arXiv:1311.2569}}].

\bibitem{Garces:2017jpz}
E.~A. Garcés, M.~Hernández~Villanueva, G.~López~Castro, and P.~Roig, {\it
  {Effective-field theory analysis of the $\tau^- \to \eta^{(\prime)} \pi^-
  \nu_\tau$ decays}},  {\it JHEP} {\bf 12} (2017) 027,
  [\href{http://arxiv.org/abs/1708.07802}{{\tt arXiv:1708.07802}}].

\bibitem{Weinberg:1978kz}
S.~Weinberg, {\it {Phenomenological Lagrangians}},  {\it Physica} {\bf A96}
  (1979), no.~1-2 327--340.

\bibitem{Gasser:1983yg}
J.~Gasser and H.~Leutwyler, {\it {Chiral Perturbation Theory to One Loop}},
  {\it Annals Phys.} {\bf 158} (1984) 142.

\bibitem{Gasser:1984gg}
J.~Gasser and H.~Leutwyler, {\it {Chiral Perturbation Theory: Expansions in the
  Mass of the Strange Quark}},  {\it Nucl. Phys.} {\bf B250} (1985) 465--516.

\bibitem{Gasser:1984ux}
J.~Gasser and H.~Leutwyler, {\it {Low-Energy Expansion of Meson Form-Factors}},
   {\it Nucl. Phys.} {\bf B250} (1985) 517--538.

\bibitem{Cata:2007ns}
O.~Cata and V.~Mateu, {\it {Chiral perturbation theory with tensor sources}},
  {\it JHEP} {\bf 09} (2007) 078, [\href{http://arxiv.org/abs/0705.2948}{{\tt
  arXiv:0705.2948}}].

\bibitem{Mateu:2007tr}
V.~Mateu and J.~Portoles, {\it {Form-factors in radiative pion decay}},  {\it
  Eur. Phys. J.} {\bf C52} (2007) 325--338,
  [\href{http://arxiv.org/abs/0706.1039}{{\tt arXiv:0706.1039}}].

\bibitem{Miranda:2018cpf}
J.~A. Miranda and P.~Roig, {\it {Effective-field theory analysis of the
  $\tau^-\to \pi^-\pi^0\nu_\tau$ decays}},
  \href{http://arxiv.org/abs/1806.09547}{{\tt arXiv:1806.09547}}.

\bibitem{Ecker:1988te}
G.~Ecker, J.~Gasser, A.~Pich, and E.~de~Rafael, {\it {The Role of Resonances in
  Chiral Perturbation Theory}},  {\it Nucl. Phys.} {\bf B321} (1989) 311--342.

\bibitem{Ecker:1989yg}
G.~Ecker, J.~Gasser, H.~Leutwyler, A.~Pich, and E.~de~Rafael, {\it {Chiral
  Lagrangians for Massive Spin 1 Fields}},  {\it Phys. Lett.} {\bf B223} (1989)
  425--432.

\bibitem{Bauer:2015knc}
M.~Bauer and M.~Neubert, {\it {Minimal Leptoquark Explanation for the
  $R_{D^{(*)}}$ , $R_K$ , and $(g-2)_\mu$ Anomalies}},  {\it Phys. Rev. Lett.}
  {\bf 116} (2016), no.~14 141802, [\href{http://arxiv.org/abs/1511.01900}{{\tt
  arXiv:1511.01900}}].

\bibitem{Gonzalez-Alonso:2016etj}
M.~González-Alonso and J.~Martin~Camalich, {\it {Global Effective-Field-Theory
  analysis of New-Physics effects in (semi)leptonic kaon decays}},  {\it JHEP}
  {\bf 12} (2016) 052, [\href{http://arxiv.org/abs/1605.07114}{{\tt
  arXiv:1605.07114}}].

\bibitem{Silva:2004gz}
J.~P. Silva, {\it {Phenomenological aspects of CP violation}},  in {\it
  {Central European School in Particle Physics Prague, Czech Republic,
  September 14-24, 2004}}, 2004.
\newblock \href{http://arxiv.org/abs/hep-ph/0410351}{{\tt hep-ph/0410351}}.

\bibitem{Sachs:1963zz}
R.~G. Sachs, {\it {Methods for Testing the CPT Theorem}},  {\it Phys. Rev.}
  {\bf 129} (1963) 2280--2285.

\bibitem{Enz:1965tr}
C.~P. Enz and R.~R. Lewis, {\it {On the phenomenological description of CP
  violation for K mesons and its consequences}},  {\it Helv. Phys. Acta} {\bf
  38} (1965) 860--876.

\bibitem{Wolfenstein:1970wb}
L.~Wolfenstein, {\it {S matrix formulation of k(l) and k(s) decays and
  unitarity relations}},  {\it Phys. Rev.} {\bf 188} (1969) 2536--2538.

\bibitem{Beuthe:1997fu}
M.~Beuthe, G.~Lopez~Castro, and J.~Pestieau, {\it {Field theory approach to K0
  - anti-K0 and B0 - anti-B0 systems}},  {\it Int. J. Mod. Phys.} {\bf A13}
  (1998) 3587--3600, [\href{http://arxiv.org/abs/hep-ph/9707369}{{\tt
  hep-ph/9707369}}].

\bibitem{AlvarezGaume:1998yr}
L.~Alvarez-Gaume, C.~Kounnas, S.~Lola, and P.~Pavlopoulos, {\it {Violation of
  time reversal invariance and CPLEAR measurements}},  {\it Phys. Lett.} {\bf
  B458} (1999) 347--354, [\href{http://arxiv.org/abs/hep-ph/9812326}{{\tt
  hep-ph/9812326}}].

\bibitem{Branco:1999fs}
G.~C. Branco, L.~Lavoura, and J.~P. Silva, {\it {CP Violation}},  {\it Int.
  Ser. Monogr. Phys.} {\bf 103} (1999) 1--536.

\bibitem{Silva:2000db}
J.~P. Silva, {\it {On the use of the reciprocal basis in neutral meson
  mixing}},  {\it Phys. Rev.} {\bf D62} (2000) 116008,
  [\href{http://arxiv.org/abs/hep-ph/0007075}{{\tt hep-ph/0007075}}].

\bibitem{Cirigliano:2018dyk}
V.~Cirigliano, A.~Falkowski, M.~González-Alonso, and A.~Rodríguez-Sánchez,
  {\it {Hadronic tau decays as New Physics probes in the LHC era}},
  \href{http://arxiv.org/abs/1809.01161}{{\tt arXiv:1809.01161}}.

\bibitem{Antonelli:2008jg}
{\bf FlaviaNet Working Group on Kaon Decays} Collaboration, M.~Antonelli
  et~al., {\it {Precision tests of the Standard Model with leptonic and
  semileptonic kaon decays}},  in {\it {PHIPSI08, proceedings of the
  International Workshop on $e^+e^-$ Collisions from phi to psi, Frascati
  (Rome) Italy, 7-10 April 2008}}, 2008.
\newblock \href{http://arxiv.org/abs/0801.1817}{{\tt arXiv:0801.1817}}.

\bibitem{Cirigliano:2009wk}
V.~Cirigliano, J.~Jenkins, and M.~Gonzalez-Alonso, {\it {Semileptonic decays of
  light quarks beyond the Standard Model}},  {\it Nucl. Phys.} {\bf B830}
  (2010) 95--115, [\href{http://arxiv.org/abs/0908.1754}{{\tt
  arXiv:0908.1754}}].

\bibitem{Bhattacharya:2011qm}
T.~Bhattacharya, V.~Cirigliano, S.~D. Cohen, A.~Filipuzzi, M.~Gonzalez-Alonso,
  M.~L. Graesser, R.~Gupta, and H.-W. Lin, {\it {Probing Novel Scalar and
  Tensor Interactions from (Ultra)Cold Neutrons to the LHC}},  {\it Phys. Rev.}
  {\bf D85} (2012) 054512, [\href{http://arxiv.org/abs/1110.6448}{{\tt
  arXiv:1110.6448}}].

\bibitem{Chang:2014iba}
H.-M. Chang, M.~González-Alonso, and J.~Martin~Camalich, {\it {Nonstandard
  Semileptonic Hyperon Decays}},  {\it Phys. Rev. Lett.} {\bf 114} (2015),
  no.~16 161802, [\href{http://arxiv.org/abs/1412.8484}{{\tt
  arXiv:1412.8484}}].

\bibitem{Buchmuller:1985jz}
W.~Buchmuller and D.~Wyler, {\it {Effective Lagrangian Analysis of New
  Interactions and Flavor Conservation}},  {\it Nucl. Phys.} {\bf B268} (1986)
  621--653.

\bibitem{Grzadkowski:2010es}
B.~Grzadkowski, M.~Iskrzynski, M.~Misiak, and J.~Rosiek, {\it {Dimension-Six
  Terms in the Standard Model Lagrangian}},  {\it JHEP} {\bf 10} (2010) 085,
  [\href{http://arxiv.org/abs/1008.4884}{{\tt arXiv:1008.4884}}].

\bibitem{Alonso:2014csa}
R.~Alonso, B.~Grinstein, and J.~Martin~Camalich, {\it {$SU(2)\times U(1)$ gauge
  invariance and the shape of new physics in rare $B$ decays}},  {\it Phys.
  Rev. Lett.} {\bf 113} (2014) 241802,
  [\href{http://arxiv.org/abs/1407.7044}{{\tt arXiv:1407.7044}}].

\bibitem{Antonelli:2010yf}
{\bf FlaviaNet Working Group on Kaon Decays} Collaboration, M.~Antonelli
  et~al., {\it {An Evaluation of $|V_{us}|$ and precise tests of the Standard
  Model from world data on leptonic and semileptonic kaon decays}},  {\it Eur.
  Phys. J.} {\bf C69} (2010) 399--424,
  [\href{http://arxiv.org/abs/1005.2323}{{\tt arXiv:1005.2323}}].

\bibitem{Moulson:2017ive}
M.~Moulson, {\it {Experimental determination of $V_{us}$ from kaon decays}},
  {\it PoS} {\bf CKM2016} (2017) 033,
  [\href{http://arxiv.org/abs/1704.04104}{{\tt arXiv:1704.04104}}].

\bibitem{Erler:2002mv}
J.~Erler, {\it {Electroweak radiative corrections to semileptonic tau decays}},
   {\it Rev. Mex. Fis.} {\bf 50} (2004) 200--202,
  [\href{http://arxiv.org/abs/hep-ph/0211345}{{\tt hep-ph/0211345}}].

\bibitem{Li:2016vvp}
X.-Q. Li, Y.-D. Yang, and X.~Zhang, {\it {Revisiting the one leptoquark
  solution to the $R(D^{(\ast)})$ anomalies and its phenomenological
  implications}},  {\it JHEP} {\bf 08} (2016) 054,
  [\href{http://arxiv.org/abs/1605.09308}{{\tt arXiv:1605.09308}}].

\bibitem{Dorsner:2016wpm}
I.~Dor\v{s}ner, S.~Fajfer, A.~Greljo, J.~F. Kamenik, and N.~Ko\v{s}nik, {\it
  {Physics of leptoquarks in precision experiments and at particle colliders}},
   \href{http://arxiv.org/abs/1603.04993}{{\tt arXiv:1603.04993}}.

\bibitem{Chetyrkin:1997dh}
K.~G. Chetyrkin, {\it {Quark mass anomalous dimension to $\mathcal
  O(\alpha_s^4)$}},  {\it Phys. Lett.} {\bf B404} (1997) 161--165,
  [\href{http://arxiv.org/abs/hep-ph/9703278}{{\tt hep-ph/9703278}}].

\bibitem{Gracey:2000am}
J.~A. Gracey, {\it {Three loop MS-bar tensor current anomalous dimension in
  QCD}},  {\it Phys. Lett.} {\bf B488} (2000) 175--181,
  [\href{http://arxiv.org/abs/hep-ph/0007171}{{\tt hep-ph/0007171}}].

\bibitem{Bernard:2011ae}
V.~Bernard, D.~R. Boito, and E.~Passemar, {\it {Dispersive representation of
  the scalar and vector $K\pi$ form factors for $\tau\to K\pi\nu_\tau$ and
  $K_{\ell3}$ decays}},  {\it Nucl. Phys. Proc. Suppl.} {\bf 218} (2011)
  140--145, [\href{http://arxiv.org/abs/1103.4855}{{\tt arXiv:1103.4855}}].

\bibitem{Brodsky:1973kr}
S.~J. Brodsky and G.~R. Farrar, {\it {Scaling Laws at Large Transverse
  Momentum}},  {\it Phys. Rev. Lett.} {\bf 31} (1973) 1153--1156.

\bibitem{Lepage:1979zb}
G.~P. Lepage and S.~J. Brodsky, {\it {Exclusive Processes in Quantum
  Chromodynamics: Evolution Equations for Hadronic Wave Functions and the
  Form-Factors of Mesons}},  {\it Phys. Lett.} {\bf 87B} (1979) 359--365.

\bibitem{Lepage:1980fj}
G.~P. Lepage and S.~J. Brodsky, {\it {Exclusive Processes in Perturbative
  Quantum Chromodynamics}},  {\it Phys. Rev.} {\bf D22} (1980) 2157.

\bibitem{Jamin:2001zr}
M.~Jamin, J.~A. Oller, and A.~Pich, {\it {Light quark masses from scalar sum
  rules}},  {\it Eur. Phys. J.} {\bf C24} (2002) 237--243,
  [\href{http://arxiv.org/abs/hep-ph/0110194}{{\tt hep-ph/0110194}}].

\bibitem{Jamin:2004re}
M.~Jamin, J.~A. Oller, and A.~Pich, {\it {Order $p^{6}$ chiral couplings from
  the scalar $K \pi$ form-factor}},  {\it JHEP} {\bf 02} (2004) 047,
  [\href{http://arxiv.org/abs/hep-ph/0401080}{{\tt hep-ph/0401080}}].

\bibitem{Coleman:1969sm}
S.~R. Coleman, J.~Wess, and B.~Zumino, {\it {Structure of phenomenological
  Lagrangians. 1.}},  {\it Phys. Rev.} {\bf 177} (1969) 2239--2247.

\bibitem{Callan:1969sn}
C.~G. Callan, Jr., S.~R. Coleman, J.~Wess, and B.~Zumino, {\it {Structure of
  phenomenological Lagrangians. 2.}},  {\it Phys. Rev.} {\bf 177} (1969)
  2247--2250.

\bibitem{Baum:2011rm}
I.~Baum, V.~Lubicz, G.~Martinelli, L.~Orifici, and S.~Simula, {\it {Matrix
  elements of the electromagnetic operator between kaon and pion states}},
  {\it Phys. Rev.} {\bf D84} (2011) 074503,
  [\href{http://arxiv.org/abs/1108.1021}{{\tt arXiv:1108.1021}}].

\bibitem{Hoferichter:2018zwu}
M.~Hoferichter, B.~Kubis, J.~Ruiz~de Elvira, and P.~Stoffer, {\it {Nucleon
  matrix elements of the tensor current}},
  \href{http://arxiv.org/abs/1811.11181}{{\tt arXiv:1811.11181}}.

\bibitem{Dekens:2018pbu}
W.~Dekens, E.~E. Jenkins, A.~V. Manohar, and P.~Stoffer, {\it {Non-Perturbative
  Effects in $\mu \to e \gamma$}},  \href{http://arxiv.org/abs/1810.05675}{{\tt
  arXiv:1810.05675}}.

\bibitem{Cata:2008zc}
O.~Cata and V.~Mateu, {\it {Novel patterns for vector mesons from the
  large-N(c) limit}},  {\it Phys. Rev.} {\bf D77} (2008) 116009,
  [\href{http://arxiv.org/abs/0801.4374}{{\tt arXiv:0801.4374}}].

\bibitem{Escribano:2013bca}
R.~Escribano, S.~Gonzalez-Solis, and P.~Roig, {\it {$\tau^-\to
  K^-\eta^{(\prime)}\nu_\tau$ decays in Chiral Perturbation Theory with
  Resonances}},  {\it JHEP} {\bf 10} (2013) 039,
  [\href{http://arxiv.org/abs/1307.7908}{{\tt arXiv:1307.7908}}].

\bibitem{Becirevic:2003pn}
D.~Becirevic, V.~Lubicz, F.~Mescia, and C.~Tarantino, {\it {Coupling of the
  light vector meson to the vector and to the tensor current}},  {\it JHEP}
  {\bf 05} (2003) 007, [\href{http://arxiv.org/abs/hep-lat/0301020}{{\tt
  hep-lat/0301020}}].

\bibitem{Donnellan:2007xr}
M.~A. Donnellan, J.~Flynn, A.~Juttner, C.~T. Sachrajda, D.~Antonio, P.~A.
  Boyle, C.~Maynard, B.~Pendleton, and R.~Tweedie, {\it {Lattice Results for
  Vector Meson Couplings and Parton Distribution Amplitudes}},  {\it PoS} {\bf
  LATTICE2007} (2007) 369, [\href{http://arxiv.org/abs/0710.0869}{{\tt
  arXiv:0710.0869}}].

\bibitem{Allton:2008pn}
{\bf RBC-UKQCD} Collaboration, C.~Allton et~al., {\it {Physical Results from
  $2+1$ Flavor Domain Wall QCD and $SU(2)$ Chiral Perturbation Theory}},  {\it
  Phys. Rev.} {\bf D78} (2008) 114509,
  [\href{http://arxiv.org/abs/0804.0473}{{\tt arXiv:0804.0473}}].

\bibitem{Cata:2009dq}
O.~Cata and V.~Mateu, {\it {Chiral corrections to the
  $f(V)^{\ast\ast}$perpendicular $/f(V)$ ratio for vector mesons}},  {\it Nucl.
  Phys.} {\bf B831} (2010) 204--216,
  [\href{http://arxiv.org/abs/0907.5422}{{\tt arXiv:0907.5422}}].

\bibitem{Dimopoulos:2011cf}
P.~Dimopoulos, F.~Mescia, and A.~Vladikas, {\it {K* vector and tensor couplings
  from Nf = 2 tmQCD}},  {\it Phys. Rev.} {\bf D84} (2011) 014505,
  [\href{http://arxiv.org/abs/1104.0188}{{\tt arXiv:1104.0188}}].

\bibitem{Guerrero:1997ku}
F.~Guerrero and A.~Pich, {\it {Effective field theory description of the pion
  form-factor}},  {\it Phys. Lett.} {\bf B412} (1997) 382--388,
  [\href{http://arxiv.org/abs/hep-ph/9707347}{{\tt hep-ph/9707347}}].

\bibitem{Pich:2001pj}
A.~Pich and J.~Portoles, {\it {The Vector form-factor of the pion from
  unitarity and analyticity: A Model independent approach}},  {\it Phys. Rev.}
  {\bf D63} (2001) 093005, [\href{http://arxiv.org/abs/hep-ph/0101194}{{\tt
  hep-ph/0101194}}].

\bibitem{Dumm:2013zh}
D.~Gómez~Dumm and P.~Roig, {\it {Dispersive representation of the pion vector
  form factor in $\tau\to\pi\pi\nu_\tau$ decays}},  {\it Eur. Phys. J.} {\bf
  C73} (2013), no.~8 2528, [\href{http://arxiv.org/abs/1301.6973}{{\tt
  arXiv:1301.6973}}].

\bibitem{Omnes:1958hv}
R.~Omnes, {\it {On the Solution of certain singular integral equations of
  quantum field theory}},  {\it Nuovo Cim.} {\bf 8} (1958) 316--326.

\bibitem{Gonzalez-Solis:2019iod}
S.~Gonzàlez-Solís and P.~Roig, {\it {A dispersive analysis of the pion vector
  form factor and $\tau ^{-}\rightarrow K^{-}K_{S}\nu _{\tau }$ decay}},  {\it
  Eur. Phys. J.} {\bf C79} (2019), no.~5 436,
  [\href{http://arxiv.org/abs/1902.02273}{{\tt arXiv:1902.02273}}].

\bibitem{Bernard:2009zm}
V.~Bernard, M.~Oertel, E.~Passemar, and J.~Stern, {\it {Dispersive
  representation and shape of the K(l3) form factors: Robustness}},  {\it Phys.
  Rev.} {\bf D80} (2009) 034034, [\href{http://arxiv.org/abs/0903.1654}{{\tt
  arXiv:0903.1654}}].

\bibitem{Hocker:2001xe}
A.~Hocker, H.~Lacker, S.~Laplace, and F.~Le~Diberder, {\it {A New approach to a
  global fit of the CKM matrix}},  {\it Eur. Phys. J.} {\bf C21} (2001)
  225--259, [\href{http://arxiv.org/abs/hep-ph/0104062}{{\tt hep-ph/0104062}}].

\bibitem{Charles:2004jd}
{\bf CKMfitter Group} Collaboration, J.~Charles, A.~Hocker, H.~Lacker,
  S.~Laplace, F.~R. Le~Diberder, J.~Malcles, J.~Ocariz, M.~Pivk, and L.~Roos,
  {\it {CP violation and the CKM matrix: Assessing the impact of the asymmetric
  $B$ factories}},  {\it Eur. Phys. J.} {\bf C41} (2005) 1--131,
  [\href{http://arxiv.org/abs/hep-ph/0406184}{{\tt hep-ph/0406184}}].

\bibitem{Li:2013vlx}
X.-Q. Li, Y.-D. Yang, and X.-B. Yuan, {\it {Exclusive radiative B-meson decays
  within minimal flavor-violating two-Higgs-doublet models}},  {\it Phys. Rev.}
  {\bf D89} (2014), no.~5 054024, [\href{http://arxiv.org/abs/1311.2786}{{\tt
  arXiv:1311.2786}}].

\bibitem{Jung:2012vu}
M.~Jung, X.-Q. Li, and A.~Pich, {\it {Exclusive radiative B-meson decays within
  the aligned two-Higgs-doublet model}},  {\it JHEP} {\bf 10} (2012) 063,
  [\href{http://arxiv.org/abs/1208.1251}{{\tt arXiv:1208.1251}}].

\bibitem{Ryu:2014vpc}
{\bf Belle} Collaboration, S.~Ryu et~al., {\it {Measurements of Branching
  Fractions of $\tau$ Lepton Decays with one or more $K^{0}_{S}$}},  {\it Phys.
  Rev.} {\bf D89} (2014), no.~7 072009,
  [\href{http://arxiv.org/abs/1402.5213}{{\tt arXiv:1402.5213}}].

\bibitem{Roig:2019rwf}
P.~Roig, {\it {Semileptonic $\tau$ decays: powerful probes of non-standard
  charged current weak interactions}},  {\it EPJ Web Conf.} {\bf 212} (2019)
  08002, [\href{http://arxiv.org/abs/1903.02682}{{\tt arXiv:1903.02682}}].

\bibitem{HFLAV:2019}
{\bf HFLAV} Collaboration, Y.~Amhis et~al., {\it {Average of $R(D)$ and
  $R(D^\ast)$ for Spring 2019}}, .
  {\url{https://hflav-eos.web.cern.ch/hflav-eos/semi/spring19/html/RDsDsstar/RDRDs.html}}.

\bibitem{Bifani:2018zmi}
S.~Bifani, S.~Descotes-Genon, A.~Romero~Vidal, and M.-H. Schune, {\it {Review
  of Lepton Universality tests in $B$ decays}},  {\it J. Phys.} {\bf G46}
  (2019), no.~2 023001, [\href{http://arxiv.org/abs/1809.06229}{{\tt
  arXiv:1809.06229}}].

\bibitem{Ciezarek:2017yzh}
G.~Ciezarek, M.~Franco~Sevilla, B.~Hamilton, R.~Kowalewski, T.~Kuhr, V.~Lüth,
  and Y.~Sato, {\it {A Challenge to Lepton Universality in B Meson Decays}},
  {\it Nature} {\bf 546} (2017) 227--233,
  [\href{http://arxiv.org/abs/1703.01766}{{\tt arXiv:1703.01766}}].

\bibitem{Li:2018lxi}
Y.~Li and C.-D. Lü, {\it {Recent Anomalies in B Physics}},  {\it Sci. Bull.}
  {\bf 63} (2018) 267--269, [\href{http://arxiv.org/abs/1808.02990}{{\tt
  arXiv:1808.02990}}].

\bibitem{Belle:2019rba}
{\bf Belle} Collaboration, G.~Caria et~al., {\it {Measurement of
  $\mathcal{R}(D)$ and $\mathcal{R}(D^*)$ with a semileptonic tagging method}},
   \href{http://arxiv.org/abs/1910.05864}{{\tt arXiv:1910.05864}}.

\end{thebibliography}\endgroup

\end{document}